\documentclass[nohyper,12pt,letterpaper]{JHEP3}
\usepackage{epsfig}
\usepackage[latin1]{inputenc}
\usepackage{bbm,amsfonts}

\author{Silvia Penati, Marco Pirrone and CarloAlberto Ratti\\
Dipartimento di Fisica, Universit\`a di Milano--Bicocca and
INFN, Sezione di Milano--Bicocca, Piazza della Scienza 3, I-20126 Milano, Italy\\
\qquad\\
E-mail: \email{silvia.penati@mib.infn.it, marco.pirrone@mib.infn.it, carloalberto.ratti@mib.infn.it}}

\abstract{We study the embedding of spacetime filling D7--branes in $\b$--deformed backgrounds 
which, according to the AdS/CFT dictionary, corresponds to flavoring $\b$--deformed ${\cal N}=4$ 
super Yang--Mills. We consider supersymmetric and more general non--supersymmetric three parameter 
deformations. The equations of motion for quadratic fluctuations of a probe D7--brane wrapped on a 
deformed three--sphere exhibit a non--trivial coupling between scalar and vector modes induced by 
the deformation. Nevertheless, we manage to solve them analytically and find that the mesonic mass 
spectrum is discrete, with a mass gap and a Zeeman--like splitting occurs. 
Finally we propose the action for the dual field theory as obtained by $\ast$--product deformation
of super Yang--Mills with fundamental matter.}

\preprint{Bicocca--FT--07--14}

\title{Mesons in marginally deformed AdS/CFT}


\csname @addtoreset\endcsname{equation}{section}

\def\bseq{\begin{subequation}}  
\def\eseq{\end{subequation}}
\def\bsea{\begin{subeqnarray}}  
\def\esea{\end{subeqnarray}}

\newcommand{\beq}{\begin{equation}}
\newcommand{\bea}{\begin{eqnarray}}
\newcommand{\eea}{\end{eqnarray}}
\newcommand{\eeq}{\end{equation}}

\newcommand {\non}{\nonumber}

\newcommand{\mc}{\mathcal}

\renewcommand{\a}{\alpha}
\renewcommand{\b}{\beta}

\renewcommand{\d}{\delta}
\newcommand{\pa}{\partial}
\newcommand{\g}{\gamma}
\newcommand{\G}{\Gamma}

\newcommand{\e}{\epsilon}

\renewcommand{\l}{\lambda}

\newcommand{\n}{\nu}

\newcommand{\F}{\Phi}

\newcommand{\s}{\sigma}
\renewcommand{\S}{\Sigma}

\renewcommand{\O}{\Omega}

\def\Mb{\kern 2pt\mathchoice
        {
         \vbox{\hrule width10pt height 0.4pt depth 0pt
         \kern 1.2pt\hbox{\kern -2pt$\displaystyle M$}}}
        {
         \vbox{\hrule width10pt height 0.4pt depth 0pt
         \kern 1.2pt\hbox{\kern -2pt$\textstyle M$}}}
        {
\vbox{\hrule width6pt height 0.4pt depth 0pt
         \kern 1.0pt\hbox{\kern -2pt$\scriptstyle M$}}}
        {
         \vbox{\hrule width5pt height 0.4pt depth 0pt
         \kern 0.8pt\hbox{\kern -2pt$\scriptscriptstyle M$}}}}

\def\Sb{\kern 2pt\mathchoice
        {
         \vbox{\hrule width6pt height 0.4pt depth 0pt
         \kern 1.2pt\hbox{\kern -2pt$\displaystyle S$}}}
        {
         \vbox{\hrule width6pt height 0.4pt depth 0pt
         \kern 1.2pt\hbox{\kern -2pt$\textstyle S$}}}
        {
         \vbox{\hrule width3.5pt height 0.4pt depth 0pt
         \kern 1.0pt\hbox{\kern -2pt$\scriptstyle S$}}}
        {
         \vbox{\hrule width3pt height 0.4pt depth 0pt
         \kern 0.8pt\hbox{\kern -2pt$\scriptscriptstyle S$}}}}

\def\Rb{\kern 2pt\mathchoice
        {
         \vbox{\hrule width5.5pt height 0.4pt depth 0pt
         \kern 1.2pt\hbox{\kern -2.5pt$\displaystyle R$}}}
        {
         \vbox{\hrule width5.5pt height 0.4pt depth 0pt
         \kern 1.2pt\hbox{\kern -2.5pt$\textstyle R$}}}
        {
         \vbox{\hrule width3.5pt height 0.4pt depth 0pt
         \kern 1.0pt\hbox{\kern -2.2pt$\scriptstyle R$}}}
        {
         \vbox{\hrule width3pt height 0.4pt depth 0pt
         \kern 0.8pt\hbox{\kern -2.2pt$\scriptscriptstyle R$}}}}

  \def\pp{{\mathchoice
      {
          \kern 1pt%
          \raise 1pt
          \vbox{\hrule width5pt height0.4pt depth0pt
            \kern -2pt
            \hbox{\kern 2.3pt
              \vrule width0.4pt height6pt depth0pt
              }
            \kern -2pt
            \hrule width5pt height0.4pt depth0pt}%
            \kern 1pt
       }
        {
          \kern 1pt%
          \raise 1pt
          \vbox{\hrule width4.3pt height0.4pt depth0pt
            \kern -1.8pt
            \hbox{\kern 1.95pt
              \vrule width0.4pt height5.4pt depth0pt
              }
            \kern -1.8pt
            \hrule width4.3pt height0.4pt depth0pt}%
            \kern 1pt
        }
        {
          \kern 0.5pt%
          \raise 1pt
          \vbox{\hrule width4.0pt height0.3pt depth0pt
            \kern -1.9pt  
            \hbox{\kern 1.85pt
              \vrule width0.3pt height5.7pt depth0pt
              }
            \kern -1.9pt
            \hrule width4.0pt height0.3pt depth0pt}%
            \kern 0.5pt
        }
        {
          \kern 0.5pt%
          \raise 1pt
          \vbox{\hrule width3.6pt height0.3pt depth0pt
            \kern -1.5pt
            \hbox{\kern 1.65pt
              \vrule width0.3pt height4.5pt depth0pt
              }
            \kern -1.5pt
            \hrule width3.6pt height0.3pt depth0pt}%
            \kern 0.5pt
        }
    }}

  \def\mm{{\mathchoice
               {
                 \kern 1pt
           \raise 1pt    \vbox{\hrule width5pt height0.4pt depth0pt
                  \kern 2pt
                  \hrule width5pt height0.4pt depth0pt}
                 \kern 1pt}
               {
                \kern 1pt
           \raise 1pt \vbox{\hrule width4.3pt height0.4pt depth0pt
                  \kern 1.8pt
                  \hrule width4.3pt height0.4pt depth0pt}
                 \kern 1pt}
               {
                \kern 0.5pt
           \raise 1pt
                \vbox{\hrule width4.0pt height0.3pt depth0pt
                  \kern 1.9pt
                  \hrule width4.0pt height0.3pt depth0pt}
                \kern 1pt}
               {
               \kern 0.5pt
         \raise 1pt  \vbox{\hrule width3.6pt height0.3pt depth0pt
                  \kern 1.5pt
                  \hrule width3.6pt height0.3pt depth0pt}
               \kern 0.5pt}
               }}

\def\pd{{\kern0.5pt
           + \kern-5.05pt \raise5.8pt\hbox{$\textstyle.$}\kern
0.5pt}}

\def\pmd{{\kern0.5pt
          \pm \kern-5.05pt
\raise6.3pt\hbox{$\textstyle.$}\kern1.5pt}}

\def\md{{\mathchoice
   {
      {{\kern 1pt - \kern-6.2pt \raise5pt\hbox{$\textstyle.$}\kern
1pt}}}
    {
      {{\kern 1pt - \kern-6.2pt \raise5pt\hbox{$\textstyle.$}\kern
1pt}}}
    {
      {\kern0.5pt - \kern-5.05pt
\raise3.4pt\hbox{$\textstyle.$}\kern0.5pt}}
    {
      {\kern0.5pt - \kern-5.05pt
\raise3.4pt\hbox{$\textstyle.$}\kern0.5pt}}}}

\def\beq{\begin{equation}}
\def\eeq{\end{equation}}
\def\bea{\begin{eqnarray}}
\def\eea{\end{eqnarray}}

\def\a{\alpha}
\def\b{\beta}
\def\g{\gamma}
\def\hg{\hat{\gamma}}
\def\d{\delta}
\def\e{\epsilon}

\def\th{\theta}
\def\l{\lambda}
\def\G{\Gamma}

\def\F{\Phi}

\begin{document}

\section{Introduction}

One of the main challenges of the elementary particle theoretical physics is the understanding
of the low energy regime of confining theories, primarily QCD. Progress in this direction is
expected in the context of AdS/CFT correspondence \cite{M} which allows for a dual description of 
Yang--Mills theories at strong coupling in terms of a perturbative string/supergravity theory.

In this respect, a quite recent progress concerns the generalization of the AdS/CFT
correspondence to include matter in the fundamental representation of the gauge group \cite{KK,KK2}. 
The holographic description of a 4D supersymmetric Yang--Mills 
theory with fundamental matter can be obtained by  considering a system of intersecting D3--D7 branes. 
Precisely, the near horizon geometry of a system of $N$ D3--branes in the presence of 
$N_f$ spacetime--filling D7--branes, in the large $N$ limit and $N_f$ fixed, gives the dual description
of a ${\cal N}=4$ $SU(N)$ SYM theory living on the D3--branes with supersymmetry broken to 
${\cal N}=2$ by $N_f$ hypermultiplets in the fundamental representation of $SU(N)$. The field
content of the hypermultiplets is given by excitations of fundamental strings stretching between
D3 and D7--branes.  

When the D3 and the D7--branes are separated along the mutual orthogonal directions the hypermultiplets
acquire a mass which is proportional to the distance between the branes. For coincident branes 
(vanishing masses) the ${\cal N}=2$ theory is superconformal invariant.  

As proposed in \cite{KK2} (see also \cite{AFM}), excitations of fundamental strings with both ends on the D7--branes
represent mesonic states of the corresponding SYM field theory. Studying these fluctuations 
allows for determining the mass spectrum of the mesonic excitations. The spectrum turns out to be
discrete with a mass gap \cite{KMMW}. 

Since the original proposal of inserting D7--branes in the standard ${\rm AdS}_5 \times {\rm S}^5$
geometry, a lot of work has been done in the direction of finding generalizations to less supersymmetric
and/or non--conformal backgrounds.
In particular, flavors and meson spectra on the conifold and in the
Klebanov--Strassler model have been studied in \cite{CKS}. The
Maldacena--Nunez background has been considered in \cite{MN}, the
class of metrics of the form ${\rm AdS}_5\times Y^{p,q}$ and ${\rm AdS}_5\times
L^{a,b,c}$ in \cite{YL}, while for the Polchinski--Strassler set--up
see \cite{PS}. Supersymmetric embeddings of D--branes and their fluctuations in 
non--commutative theories have been investigated in \cite{NC}. Further
generalizations concern other stable brane systems
\cite{defect,defect2}. Chiral symmetry breaking and theories at
finite temperature have been first studied in
\cite{chiral1,chiral2}. Moreover, several attempts have been devoted
to going beyond the probe approximation and studying full back--reacted
(super)gravity solutions \cite{back}. Further interesting results
can be found in \cite{kirsch,other2,other3,other4,other5}.

Among the formulations of the AdS/CFT correspondence with less
supersymmetry, the one--parameter Lunin--Maldacena (LM)
background \cite{LM} corresponding to ${\cal N}=1$ $\beta$--deformed
SYM theories plays an interesting role, being the field theory and the
dual string geometry explicitly known.  The gravitational background
is ${\rm AdS}_5 \times \tilde{{\rm S}}^5$ where $\tilde{{\rm S}}^5$ is
the $\beta$--deformed five sphere obtained by performing a $TsT$
transformation on a 2--torus inside the ${\rm S}^5$ of the
original background.  This operation breaks the $SO(6)$ symmetry group
of the five sphere down to $U(1) \times U(1) \times U(1)$.  On the
field theory side, this deformation corresponds to promoting the
ordinary products among the fields in the ${\cal N}=4$ action to a
$\ast$--product which depends on the charges of the fields under two 
$U(1)$'s and allowing for the chiral coupling constant to be
different from the gauge coupling. Consistently with what happens on
the string side, these operations break ${\cal N}=4$ to ${\cal N}=1$
supersymmetry, as the third $U(1)$ (the one not involved in the 
$\ast$--product) corresponds to the R--symmetry.  Further generalizations
\cite{Frolov} lead to a dual correspondence between a
non--supersymmetric Yang--Mills theory and a deformed LM background
depending on three different real parameters $\g_1$, $\g_2$ and
$\g_3$ \footnote{We use the standard convention to name 
{\em real} deformation parameters with $\g$.}.

All these models are (super)conformal invariant since the string
geometry still has an AdS factor. As such they cannot be used to give
a realistic description of the RG flow of a gauge theory towards a
confining phase.  However, it is interesting to investigate
what happens if we insert D7--branes in these deformed
backgrounds \footnote{Several works in the literature are devoted to the
study of D--branes in this context \cite{dbef,hsz,STV,GODB,Ni,mariotti,Klu}.}. 
In particular, we expect to find a
parametric dependence of the mesonic spectrum on $\g_i$'s
which could then be used to fine--tune the results.

In what follows we accomplish this project by studying the effects 
of inserting D7--branes in the more general non--supersymmetric LM--Frolov
background. In the probe approximation ($N_f \ll N$), we first study
the stability of the D3--D7 configuration. We find that, independently
of the value of the deformation parameters, an embedding can be found
which is stable, BPS and in the $\g_1=\g_2=\g_3$ case it is also
supersymmetric. 

We then study fluctuations of a D7--brane around the static embedding which correspond 
to scalar and vector mesons of the dual field theory. 
We consider the equations of motion for the tower of Kaluza--Klein
modes arising from the compactification of the D7--brane on a deformed three--sphere.
The background deformation induces a non--trivial coupling between scalar and vector modes.
However, with a suitable field redefinition, we manage to simplify the equations
and solve them analytically, so determining the mass spectrum exactly.

The effects of the deformation on the mesonic mass spectrum and on the corresponding KK modes
are the following:
i) As in the undeformed case the mass spectrum is discrete and with a mass gap, but
it acquires a non--trivial dependence on the deformation parameters. Precisely, it
depends on the parameters $\g_2, \g_3$ which are associated to $TsT$ transformations
along the tori with a direction orthogonal to the probe branes, whereas the parameter $\g_1$ 
associated to the
deformation along the torus inside the D7 worldvolume never enters the equations of motion for
quadratic fluctuations and does not affect the mass spectrum. 
ii) Since the deformation breaks $SO(4)$ (the isomorphisms of the three--sphere)
to $U(1) \times U(1)$ a Zeeman--like effect occurs and the masses exhibit a non--trivial dependence 
on the $(m_2,m_3)$ quantum numbers associated to the two $U(1)$'s. 
The dependence is through the linear combination $(\g_2 m_3 - \g_3 m_2)^2$ 
so that the mass eigenvalues are smoothly related to the ones of the undeformed case by sending $\g_i \to 0$.
iii) The corresponding eigenstates are classified according to their $SO(4)$ 
and $U(1) \times U(1)$ quantum numbers. Expanding in vector and scalar harmonics on the 
three--sphere, we find  Type I elementary fluctuations
\footnote{We use the classification of \cite{KMMW}.} in the  
$(\frac{l\mp 1}{2}, \frac{l\pm 1}{2})_{(m_2,m_3)}$ representations and Type II, Type III and scalar modes in the
$(\frac{l}{2}, \frac{l}{2})_{(m_2,m_3)}$.
For a given $l$ the total number of degrees of freedom is $8(l+1)^2$ as in the undeformed theory but, given the
degeneracy breaking, they split among different eigenvalues. For 
any given triplet $(l,m_2,m_3)$ we compute the degeneracy of the corresponding mass eigenvalue. We find that
the splitting is different according to the choice $\g_2 \neq \g_3$ or $\g_2=\g_3$ (which
includes the $\mathcal{N}=1$ supersymmetric deformation). In the last case the spectrum exhibits a mass degeneracy 
between scalars and vectors which is remnant of the $\mathcal{N}=2$ supersymmetric, undeformed case. 
 
The paper is organized as follows. In Section 2 we review the three--parameter deformation of the 
${\rm AdS}_5\times {\rm S}^5$ by using a set of coordinates suitable for the introduction of D7--branes. 
In Section 3 we study the static embedding of a D7--brane and discuss its stability. 
In the $\g_1=\g_2=\g_3$ case, using the results of \cite{mariotti}  we argue that our configuration 
is supersymmetric. 
We then find the equations of motion for the bosonic fluctuations of a D7--brane in Section 4 and 
solve them analytically in Section 5 determining the exact mass spectrum. In Section 6 we 
discuss the properties of the spectrum and analyze in detail the splitting of the mass levels and 
the corresponding degeneracy. Finally, in Section 7 we 
formulate the field theory dual to our configuration, whereas our conclusions, comments and 
perspectives are collected in Section 8.

\section{Generalities on the three--parameter deformation of ${\rm AdS}_5\times {\rm S}^5$}
 
Following \cite{LM, Frolov} we consider a type IIB supergravity background obtained as a 
three--parameter 
deformation of ${\rm AdS}_5\times {\rm S}^5$. It is realized by three $TsT$ transformations 
(T duality -- angle shift -- T duality) 
along three tori inside ${\rm S}^5$ and driven by three different real parameters $\g_i$.
The corresponding metric is usually written in terms of radial/toroidal coordinates $(\rho_i, \phi_i)$, 
$i=1,2,3$, $\sum_i \rho_i^2 = 1$ on the deformed sphere, and in string frame it reads 
(we set $\alpha'=1$)
\bea\label{metric0}
&& ds^2 = \frac{u^2}{R^2} \eta_{\mu \nu} dx^\mu dx^\nu + \frac{R^2}{u^2} du^2 +
R^2 \left[ \sum_i (d\rho_i^2 + G \rho_i^2 d\phi_i^2) + G \rho_1^2 \rho_2^2 \rho_3^2 \left(
\sum_i \hat{\gamma}_i d\phi_i \right)^2 \right]
\nonumber \\
&& G^{-1} = 1 + \hat{\gamma}_3^2 \rho_1^2 \rho_2^2 + \hat{\gamma}_2^2 \rho_3^2 \rho_1^2 + 
\hat{\gamma}_1^2 \rho_2^2 \rho_3^2 \qquad \qquad \qquad \hat{\gamma}_i \equiv R^2 \gamma_i  
\eea
where $R$ is the ${\rm AdS}_5$ and ${\rm S}^5$ radius. A further change of coordinates may be useful 
(we use the notation $c_\xi \equiv \cos{\xi}, s_\xi\ \equiv \sin{\xi}$ for any angle $\xi$)
\beq
\rho_1 = c_{\alpha} \quad , \quad  \rho_2 = s_{\alpha} c_{\theta} \quad , \quad 
\rho_3 = s_{\alpha} s_{\theta} 
\eeq
leading to the description of this background in terms of Minkowski coordinates $x^\mu$ plus the ${\rm AdS}_5$
coordinate $u$ and five angular coordinates $(\alpha, \theta, \phi_1, \phi_2, \phi_3)$.
The deformations correspond to $TsT$ transformations along the three tori $(\phi_1,\phi_2)$, $(\phi_1,\phi_3)$, 
$(\phi_2,\phi_3)$ and are parametrized by constants $\hg_3$, $\hg_2$ and $\hg_1$ respectively. 

This background is non--supersymmetric and it is dual
to a non--supersymmetric but marginal deformation of $\mathcal{N}=4$ SYM (the deformation has to be exactly
marginal since the AdS factor is not affected by $TsT$'s). The ${\cal N} =1$ supersymmetric background 
of \cite{LM} can be recovered by setting $\hg_1 = \hg_2 =\hg_3$.

With the aim of embedding D7--branes in this background we find  more convenient to express the 
metric in terms of a slightly different set of coordinates. We describe the six dimensional
internal space in terms of 
$X^m \equiv \{\rho,\theta,\phi_2,\phi_3,X_5,X_6\}$ which are mapped into the previous set of 
coordinates by the change of variables
\beq
\rho=u\, s_\alpha \quad ,\quad X_5=u\, c_\alpha\, c_{\phi_1} \quad ,\quad  X_6=u\, c_\alpha\, s_{\phi_1}
\label{change}
\eeq
In string frame and still setting $\alpha'=1$, we then have
\beq\label{metric}
ds^2=\frac{u^2}{R^2}\,\eta_{\mu \nu} dx^\mu dx^\nu +\frac{R^2}{u^2}\,G_{mn}dX^m dX^n
\eeq
where the non--vanishing components of the metric $G_{mn}$ are
\bea\label{metric2}
G_{\rho\rho}&=& \,1 \qquad \qquad 
G_{\theta\theta}=\, \rho^2 \non \\
&&\non \\
G_{\phi_2\phi_2}&=& G \left(1+\hat{\gamma}_2^2 \rho_1^2 \rho_3^2\right) \rho_2^2 \, u^2 
\qquad \quad\qquad\quad\,\,\,\,\,
G_{\phi_3\phi_3}= G \left(1+\hat{\gamma}_3^2 \rho_1^2 \rho_2^2\right) \rho_3^2 \, u^2 \non\\ 
&~&~~~~~~~~~~~~~~~~~~~~G_{\phi_2\phi_3}= G \,\hat{\gamma}_2 \hat{\gamma}_3\,  
\rho_1^2 \rho_2^2 \rho_3^2\, u^2 \non \\
&&\non \\
G_{\phi_2 X_5}&=&-G\,\hat{\gamma}_1\hat{\gamma}_2 \, \rho_2^2 \rho_3^2 \,X_6 \qquad \qquad
\qquad \qquad\,\,
G_{\phi_2 X_6} = G\,\hat{\gamma}_1\hat{\gamma}_2 \, \rho_2^2 \rho_3^2 \,X_5 \non\\
G_{\phi_3 X_5}&=& -G\,\hat{\gamma}_1\hat{\gamma}_3 \, \rho_2^2 \rho_3^2 \,X_6\qquad \qquad\,\,
\qquad \qquad
G_{\phi_3 X_6} =G\,\hat{\gamma}_1\hat{\gamma}_3 \, \rho_2^2 \rho_3^2 \,X_5  \non\\
&&\non \\
G_{X_5 X_5}&=& 1- \frac{X_6^2}{u^2 \rho_1^2}\left[1-G \left(1+\hat{\gamma}_1^2 \rho_2^2 
\rho_3^2\right)\right]\qquad
G_{X_6 X_6} = 1-\frac{X_5^2}{u^2 \rho_1^2}\left[1-G \left(1+\hat{\gamma}_1^2 \rho_2^2 
\rho_3^2\right)\right] \non\\
&~&~~~~~~~~~~~~G_{X_5 X_6}= \frac{X_5 X_6}{u^2 \rho_1^2}\left[1-G \left(1+\hat{\gamma}_1^2 
\rho_2^2 \rho_3^2\right)\right]
\eea
where $G$ is given in (\ref{metric0}) and now 
\beq
\rho_1^2=\frac{X_5^2+X_6^2}{u^2} \quad , \quad
\rho_2^2=\frac{\rho^2 c_\theta^2}{u^2} \quad , \quad
\rho_3^2=\frac{\rho^2 s_\theta^2}{u^2}
\eeq
The constraint $\sum_{i=1}^3 \rho_i^2=1$ is traded with the condition 
$u^2=\rho^2+X_5^2+X_6^2$.

The LM--Frolov supergravity solution is characterized by a non--constant dilaton
\beq
e^{2\phi}=e^{2\phi_0}G
\label{dilaton}
\eeq
where $\phi_0$ is the constant dilaton of the undeformed background
related to the AdS radius by $R^4=4\pi e^{\phi_0}N \equiv \lambda$.
For real deformation parameters $\hat{\gamma}_i$ the axion field $C_0$ is a 
constant and can be set to zero.

This background carries also a non--vanishing NS-NS two--form and R-R forms as well.
In our set of coordinates they read 
\bea\label{B}
B &=& \frac{R^2 G}{u^2} \Bigg((X_5 dX_6-X_6 dX_5) \wedge 
(\hat{\gamma}_3 \rho_2^2 d\phi_2-\hat{\gamma}_2 \rho_3^2 d\phi_3)+
\hat{\gamma}_1\rho_2^2\rho_3^2\,u^2 d\phi_2\wedge d\phi_3 \Bigg)
\non \\
C_2 &=& 4 R^2 e^{-\phi_0} \omega_1 \wedge\left(\hat{\gamma}_1\frac{X_5 dX_6-X_6 dX_5}{u^2 \rho_1^2}+
\hat{\gamma}_2 d\phi_2+\hat{\gamma}_3 d\phi_3\right)~\, ,
\quad\quad \omega_1=\frac{\rho^4}{4 u^4}c_\theta s_\theta d\theta
\non \\
C_4 &=& 4 R^4 e^{-\phi_0}\left(\frac{u^4}{4 R^8}dt\wedge dx_1\wedge dx_2 \wedge dx_3\ - 
G\, \omega_1 \wedge\frac{X_5 dX_6-X_6 dX_5}{u^2 \rho_1^2} \wedge d\phi_2 \wedge d\phi_3\right)
\non\\
&&~~
\eea
The corresponding field strengths are given by the general prescription 
$\tilde{F}_q=dC_{q-1}-dB\wedge C_{q-3}$. 

The missing forms of higher degrees can be found by applying the ten--dimensional Hodge 
duality operator
\beq\label{duals}
\tilde{F}_7=-\star \tilde{F}_3,\qquad\qquad
\tilde{F}_9=\star \tilde{F}_1 
\eeq
From the first identity and using the equation of motion for $C_2$ 
\beq
d(\star \tilde{F}_3)=dC_4\wedge dB,
\eeq
it is easy to see that $d(C_6-B \wedge C_4)=0$, i.e. $C_6-B \wedge C_4=dX$ for an arbitrary 5--form $X$.
We make the gauge choice
\beq
C_6=C_4\wedge B
\label{C6}
\eeq 
Finally, from the second identity in (\ref{duals}), by using (\ref{C6}) and taking into account that
$B \wedge B = 0$ and $C_0 = 0$ we find $\tilde{F}_9=dC_8=0$. Therefore, in what follows we set $C_8=0$.

The deformed background written in terms of the original internal coordinates 
$(\rho, \alpha, \theta, \phi_1, \phi_2, \phi_3)$ has a manifest invariance under constant shifts
of the toroidal coordinates $(\phi_1, \phi_2, \phi_3)$ which correspond to three $U(1)$ symmetries. 
With our choice of coordinates the invariance under $\phi_{2,3} \to \phi_{2,3} + {\rm const.}$ is
still manifest, whereas the third $U(1)$ associated to shifts of $\phi_1$ is realized as a rotation in the
$(X_5,X_6)$ plane.

\section{The embedding of D7--branes}

We now study the embedding of $N_f \ll N$ D7--branes in the deformed background described in the previous 
Section. For simplicity we consider the case of a single spacetime filling D7--brane ($N_f =1$)
which extends in the internal directions $(\rho, \theta, \phi_2, \phi_3)$ (we work in the 
static gauge where the worldvolume coordinates $\sigma^a$ of the brane are identified with the appropriate 
ten dimensional coordinates). 
The $X_5,X_6$ coordinates parametrize the mutual orthogonal directions of the intersecting 
system of $N$ sources D3--branes and one flavor D7--brane.

The dynamics of bosonic degrees of freedom of the D7--brane is described by the action
\beq\label{dbiwz}
S=S_{DBI}+S_{WZ}
\eeq
where $S_{DBI}$ is the abelian Dirac--Born--Infeld term (in what follows 
latin labels $a,b,...$ stand for worldvolume components)
\beq\label{sdbi}
S_{DBI}=-T_7\int_{\Sigma_8} d^8\sigma\, e^{-\phi}\sqrt{-det(g_{ab}+\mathcal{F}_{ab})}
\eeq
whereas $S_{WZ}$ is
the Wess--Zumino term describing the coupling of the brane to the R-R potentials
\bea\label{Swz}
S_{WZ}&=&T_7\int_{\Sigma_8}\left\{\frac{(2\pi\alpha')^3}{6}P[C_2]\wedge F\wedge F\wedge F+
\frac{(2\pi\alpha')^2}{2}P[C_4-C_2\wedge B]\wedge F\wedge F\right\}
\non \\
&&~~~
\eea
Here $g_{ab} \equiv G_{MN}\partial_a X^M \partial_b X^N$ is the pull--back
of the ten--dimensional spacetime metric (\ref{metric}, \ref{metric2}) on the 
worldvolume $\Sigma_8$ and $T_7$ is the D7--brane tension.
The $U(1)$ worldvolume gauge field strength $F_{ab}$ enters the action through
the modified field strength $\mathcal{F}_{ab}=2\pi \alpha' F_{ab}-b_{ab}$,
where $b_{ab}$ is the pull--back of the target NS-NS two--form potential in (\ref{B}), 
$b_{ab}=B_{MN}\partial_a X^M \partial_b X^N$. Moreover, in (\ref{Swz}) 
$P[...]$ denotes the pull--back of the R-R forms on $\Sigma_8$.

\vskip 15pt
We look for ground state configurations of the D7--brane. These are static solutions of the 
equations of motion for $X_5, X_6$ and $\varepsilon F$
($\varepsilon \equiv 2\pi\alpha'$) derived from (\ref{dbiwz}). 

In the ordinary ${\rm AdS}_5 \times {\rm S}^5$ background static embeddings 
(see for example \cite{chiral1})
can be found by setting $X_6=0 $, $F=0$ and $X_5= X_5(\rho)$ satisfying 
\beq\label{X5eq}
\frac{d}{d\rho}\left(\frac{\rho^3}{\sqrt{1+(\partial_\rho X_5)^2}}\frac{d X_5}{d\rho}\right)=0
\eeq
with asymptotic behavior $X_5(\rho)=L+\frac{c}{\rho^2}$ for $\rho\gg 1$.
The mass solution $X_5=L$ is the only well--behaved solution and corresponds
to fixing the location of the D7--brane in the 56--plane at $X_5^2+X_6^2=L^2$. This is a 
BPS configuration since the energy density turns out to be independent of $L$ \cite{Noforce,defect2}.

In the deformed background we consider an embedding of the form
\beq\label{embe}
X^M =(x_\mu,\rho,\theta,\phi_2,\phi_3,X_5(\rho),X_6(\rho)) 
\quad , \quad F = F(X^M)
\eeq
where, as in the ordinary case, we allow for a non--trivial dependence of the 
orthogonal directions on the non--compact internal coordinate $\rho$. 
Solving the equations of motion for $X_5,X_6$ and $F$ in the present case requires 
a bit of care since the non--vanishing NS-NS 2--form in (\ref{B}) 
can act as a source for the field strength $\varepsilon F$. 

We expand the action (\ref{dbiwz}) up to second order in $\varepsilon F$.
The WZ action is simply 
\beq\label{LWZ}
S_{WZ}=\frac{T_7}{2} \, \int_{\S_8} P[C_4-C_2\wedge B]\wedge \varepsilon F
\wedge \varepsilon F
\eeq
whereas the expansion of $S_{DBI}$ gives 
\bea \label{kasun}
\mathcal{L}_{DBI} &=&-T_7\frac{\sqrt{-det{\left(g-b+\varepsilon F\right)}}}{\sqrt{G}} \non\\
&=&-T_7\frac{\sqrt{-det{\left(g-b\right)}}}{\sqrt{G}}\sqrt{det{\left(1+Y\right)}} \nonumber \\
&=&-T_7 \, \rho^3 s_\theta c_\theta\,\sqrt{\Omega_2}\,\left\{1+\frac{1}{2}\mbox{Tr}(Y)
-\frac{1}{4}\mbox{Tr}(Y^2)
+\frac{1}{8}\left[\mbox{Tr}(Y)\right]^2+\cdots\right\}
\eea
where we have defined
\bea\label{Y}
Y &\equiv& \left(g-b\right)^{-1} \varepsilon F
\non\\
\Omega_2 &\equiv& 1+(\partial_\rho X_5)^2+(\partial_\rho X_6)^2
\eea
and set $e^{\phi_0} \equiv 1$.

The source for $\varepsilon F$ comes from the term 
\beq
\frac12 \mbox{Tr}(Y) = \frac{\varepsilon}{R^2 \Omega_2}\,
\left[(X_5\,\partial_\rho X_6-X_6\,\partial_\rho X_5)(\hat{\gamma}_2\,F_{\rho\phi_3}-\hat{\gamma}_3
\,F_{\rho\phi_2})
-\hat{\gamma}_1\Omega_2\,F_{\phi_2\phi_3}\right]
\eeq
In the abelian case the last term is a total derivative and, once integrated on the 
worldvolume of the brane, it cancels. We are left with the first term which gives 
a non--trivial coupling between the scalars and the vectors. 
We note that these couplings are proportional to the deformation
parameters and disappear for $\hat{\g}_i=0$, consistently with the undeformed case.

Since all the $F$ components except $F_{\rho\phi_2}$ and $F_{\rho\phi_3}$ satisfy homogeneous 
equations we can set them to zero and concentrate on the system of coupled equations of motion 
for $X_5, X_6, F_{\rho\phi_2}$ and $F_{\rho\phi_3}$. 
It is easy to realize that a solution is still given by $X_6 =0$, $F_{\rho\phi_2}= F_{\rho\phi_3}=0$,
whereas $X_5(\rho)$ satisfies eq. (\ref{X5eq}) and can be chosen as $X_5 = L$. 

Therefore, even in the deformed case, the ground state of the probe brane is given by 
a static location at $X_5^2 + X_6^2 = L^2$ with no $F$ flux and absence of non--trivial quark
condensate. The choice $X_5=L$ and $X_6=0$ 
breaks the rotational invariance in the $(X_5,X_6)$ plane. 
   
This configuration is stable (BPS). In fact, the corresponding action 
\beq\label{stable}
S = -T_7 \int_{\S_8} d^8 \s \rho^3 s_\theta c_\theta 
\eeq
coincides with the one of the undeformed case and 
satisfies the no--force condition \cite{Noforce,defect2}.

Setting $X_5^2 +X_6^2 = L^2$, the induced metric on the D7--brane reads 
\bea\label{metrind}
ds_I^2&\equiv& g_{ab}\, dX^a dX^b 
\non\\
&=& \frac{L^2+\rho^2}{R^2}\left(-dt^2+dx_1^2+dx_2^2+dx_3^2\right)+
\frac{R^2}{L^2+\rho^2}(d\rho^2+\rho^2 d\theta^2)\non\\
&+&\frac{R^2 G \rho^2}{(L^2+\rho^2)}\left[c_\theta^2 d\phi_2^2+s_\theta^2 d\phi_3^2+
\frac{\rho^2 L^2 c_\theta^2 s_\theta^2 (\hat{\gamma}_2 d\phi_2+\hat{\gamma}_3 
d\phi_3)^2}{(L^2+\rho^2)^2}\right]
\eea  
where $G$ in (\ref{metric0}) takes the explicit form
\beq\label{Gemb}
G=\frac{(L^2+\rho^2)^2}{(L^2+\rho^2)^2+
\hat{\gamma}_1^2 \rho^4 s_\theta^2 c_\theta^2+
\hat{\gamma}_2^2 L^2 \rho^2 s_\theta^2+
\hat{\gamma}_3^2 L^2 \rho^2 c_\theta^2}
\eeq

We note that, due to the particular embedding we have realized, the parameter
$\hat{\gamma}_1$ associated to the $TsT$ transformation on the $(\phi_2, \phi_3)$ torus
inside the D7 worldvolume enters the metric differently from $\hat{\gamma}_{2,3}$ which are
instead associated to deformations on tori with one parallel and one orthogonal direction 
to the probe. 

The different role played by $\hat{\gamma}_1$ respect to $(\hat{\gamma}_2, \hat{\gamma}_3)$ 
can be also understood by looking at the conformal case ($L=0$) or the UV limit ($\rho \to \infty$)
of the theory. In  both cases the dependence on $(\hat{\gamma}_2, \hat{\gamma}_3)$ disappears and
the worldvolume metric reduces to the one for ${\rm AdS}_5 \times \tilde{{\rm S}}^3$
where $\tilde{{\rm S}}^3$ is the deformed three--sphere with metric
\beq\label{S3def}
\frac{ds^2_{\tilde{{\rm S}}^3}}{R^2} = d\theta^2 + G ( c_\theta^2 d\phi_2^2 + s_\theta^2
d\phi_3^2) \quad , \quad 
G = \frac{1}{1+ \hat{\gamma}_1^2 c_\theta^2 s_\theta^2}
\eeq
Instead, for $\rho$ finite and $L \neq 0$ the ${\rm AdS}_5$ factor is lost, the theory is no longer
conformal and a non--trivial dependence on all the deformation parameters appears.

The particular probe brane configuration we have chosen is 
smoothly related to the one of the undeformed case. 
In fact, sending $\hg_i \to 0$ 
we recover the usual Karch--Katz \cite{KK} picture of flavor branes in 
${\rm  AdS}_5\times {\rm S}^5$. As we have just proved,
the stability of the D3--D7 system survives the deformation. 

We have embedded flavor D7--branes in a deformed background. When the D7--brane 
is spacetime filling and wraps the $(\phi_2, \phi_3)$ torus the configuration
is stable and no worldvolume flux is turned on. 
Alternatively, we could have started with a configuration of D7--branes
in the undeformed ${\rm  AdS}_5\times {\rm S}^5$ background and perform the three 
$TsT$ transformations as a second step. If the D7--branes were to be placed along the
same directions as before, we would obtain exactly the same configuration
of stable D7--branes in the deformed background with no flux turned on. 
In fact, along the directions $(\phi_1,\phi_2, \phi_3)$ affected by $TsT$
transformations the probe branes have Dirichlet--Neumann--Neumann (DNN) boundary 
conditions. Considering the proposal in \cite{STV} and according to the analysis 
of \cite{Ni} a DNN configuration
with no flux is mapped into the same configuration, whatever is
the $TsT$ transformation we perform. 
Therefore, for the particular embedding we are analyzing the two operations
i) Adding a probe to the deformed background and
ii) Performing a $TsT$ transformation on the undeformed brane scenario
are equivalent processes.
The stability of our brane configuration for any value of the deformation parameters
then follows from the fact that $TsT$ transformations do not affect 
the BPS nature of the original brane system \cite{LM} (see also \cite{GODB}). 

It is worth stressing that the possibility of applying equivalently 
prescriptions i) or ii) 
is peculiar of the particular brane configuration
we have chosen. Had we considered different embeddings, the two 
procedures wouldn't had led necessarily to equivalent settings \cite{STV,Ni}. 
Furthermore, the stability of the configuration would have become questionable. 

When the deformation parameters $\hat{\gamma}_i$ are all equal the 
${\rm AdS}_5 \times \tilde{{\rm S}}^5$ background has ${\cal N}=1$ supersymmetry. 
The question is whether our D7--brane embedding preserves 
supersymmetry. 
The standard way of finding supersymmetric configurations is  to look at
the  $\kappa$--symmetry  condition  of  the  probes. However, since  
the $\b$--deformed background can be described by an $SU(2)$ structure manifold,
it is  more convenient  to work using  the formalism  of G--structures
\cite{Gau} and Generalized Complex Geometry (GCG) \cite{MPZ}.
In this framework the supersymmetry conditions for D--branes
probing $SU(2)$ structure manifolds have been established in \cite{mariotti}.
For spacetime filling D7--branes a class of supersymmetric embeddings is
given by $z_1 \equiv X_5+i X_6=L$, with $z_2 \equiv X_1+i X_2$ and $z_3 \equiv
X_3+i X_4$ arbitrarily fixed and no worldvolume flux turned on.
This embeddings break one of the $U(1)$ global symmetries. 
Since our configuration belongs to this class we conclude that our 
embedding is supersymmetric.

\section{Probe fluctuations}

As proposed in \cite{KK2,AFM} D7--brane fluctuations around its ground state are 
dual to color singlets which may be interpreted as describing mesonic
states of the four dimensional gauge theory. The mass spectrum of the mesons is 
given by the Kaluza--Klein spectrum of states which originate from the compactification
of the D7--brane on the internal submanifold. In the ordinary undeformed scenario 
the spectrum is discrete and with a mass gap \cite{KMMW}.  

Our main purpose is to investigate probe fluctuations in the deformed background.

A generic vibration of the brane around its ground state can be described by 
\beq
X_5=L+\varepsilon \,\chi(\sigma^a)\,,\qquad\qquad X_6=\varepsilon \,\varphi(\sigma^a)
\eeq
together with a non--trivial flux $\varepsilon F_{ab} = \varepsilon (\pa_a A_b - \pa_b A_a)$. 
The fluctuations are functions of the worldvolume coordinates $\sigma^a$ 
and $\varepsilon$ is a small perturbation parameter. 

We expand the action of the probe brane in powers of the small parameter 
\beq
S= S_{DBI} + S_{WZ} = 
\int_{\S_8} d^8\sigma\{\mathcal{L}_0+\varepsilon \mathcal{L}_1+\varepsilon^2\mathcal{L}_2+\cdots\}
\eeq
and consider terms up to the quadratic order in $\varepsilon$.

We first concentrate on the DBI term 
\beq\label{lag}
\mathcal{L}_{DBI}=-T_7\frac{1}{\sqrt{G}}\sqrt{-det(g-b+\varepsilon F)}
\eeq
where we have written the dilaton field as in (\ref{dilaton}) with $e^{\phi_0} \equiv 1$.

We expand the various terms by writing
\bea
&& g=g^{(0)}+\varepsilon g^{(1)}+\varepsilon^2 g^{(2)}\,,
\qquad b=b^{(0)}+\varepsilon b^{(1)}+\varepsilon^2 b^{(2)}
\non \\
&& \frac{1}{\sqrt{G}}=G^{(0)}+\varepsilon G^{(1)}+\varepsilon^2 G^{(2)} 
\eea
Therefore, the determinant can be written as
\bea
\sqrt{-det(g-b+\varepsilon F)}&=&\sqrt{-det\left(g^{(0)}-b^{(0)}\right)}\sqrt{det(1+Y)}
\non \\
&=& \sqrt{-det\left(g^{(0)}-b^{(0)}\right)} \left[1+\frac{1}{2}\mbox{Tr}(Y)-\frac{1}{4}\mbox{Tr}(Y^2)+
\frac{1}{8}\left[\mbox{Tr}(Y)\right]^2+\cdots \right]
\non \\
&&~~~~
\label{Yexp}
\eea
where the matrix $Y$ is given by
\beq
Y = \left(g^{(0)}-b^{(0)}\right)^{-1}\left[\varepsilon\left(g^{(1)}-b^{(1)}+F\right)+
\varepsilon^2\left(g^{(2)}-b^{(2)}\right) + \cdots \right]
\label{Yexp2}
\eeq
At the lowest order the contribution $g^{(0)}$ is easily read from (\ref{metrind}), whereas for the
pull--back of $B$ from eq. (\ref{B}) we find that the only non--vanishing component is
$b^{(0)}_{\phi_2\phi_3} = \hat{\gamma}_1 R^2 G \rho_2^2 \rho_3^2$.

It is convenient to introduce the undeformed induced metric 
\bea\label{auxmet}
\mathcal{G}=diag\left(-\frac{L^2+\rho^2}{R^2},\frac{L^2+\rho^2}{R^2},
\frac{L^2+\rho^2}{R^2},\frac{L^2+\rho^2}{R^2},
\frac{R^2}{L^2+\rho^2},\frac{R^2 \rho^2}{L^2+\rho^2},
\frac{R^2 \rho^2 c^2_\theta}{L^2+\rho^2},\frac{R^2 \rho^2 s^2_\theta}{L^2+\rho^2}\right)\non\\
\eea
the auxiliary metric $\mathcal{C}$ defined by 
\bea
d\hat{s}^2&\equiv& \mathcal{C}_{ab}d\sigma^ad\sigma^b
\non \\
&=& \frac{L^2+\rho^2}{R^2}\left(-dt^2+dx_1^2+dx_2^2+dx_3^2\right)+\frac{R^2}{L^2+\rho^2}(d\rho^2+
\rho^2 d\theta^2)\non\\
&+&\frac{R^2 \hat{G} \rho^2}{L^2+\rho^2}\left[c_\theta^2 d\phi_2^2+s_\theta^2 d\phi_3^2+
\frac{\rho^2 L^2 c_\theta^2 s_\theta^2 (\hat{\gamma}_2 d\phi_2+\hat{\gamma}_3 d\phi_3)^2}
{(L^2+\rho^2)^2}\right]
\eea  
with
\beq
\hat{G} = \frac{(L^2+\rho^2)^2}{(L^2+\rho^2)^2+\hat{\gamma}_2^2 L^2 \rho^2 s_\theta^2+
\hat{\gamma}_3^2 L^2 \rho^2 c_\theta^2}
\eeq
and two deformation matrices ${\cal T}$ and ${\cal J}$ given by
\bea\label{T}
&&\mathcal{T}^{\phi_2\phi_2}=\hat{\gamma}_3^2\,\qquad\,\, \mathcal{T}^{\phi_3\phi_3}=\hat{\gamma}_2^2
\, \qquad\,\, \mathcal{T}^{\phi_2\phi_3}=\mathcal{T}^{\phi_3\phi_2}=-\hat{\gamma}_2\hat{\gamma}_3
\non\\
&&\mathcal{J}^{\phi_2\phi_2}= 0 \, \qquad\,\, \mathcal{J}^{\phi_3\phi_3} \,= 0 \, \qquad\,\,
\, \, \mathcal{J}^{\phi_2\phi_3}=-\mathcal{J}^{\phi_3\phi_2}=\g_1
\eea
The metric $\mathcal{C}$ is nothing but the induced metric (\ref{metrind}) evaluated at $\hat{\gamma}_1=0$.
Its inverse can be expressed as 
\beq\label{Cinv}
\mathcal{C}^{-1}=\mathcal{G}^{-1}+\frac{L^2}{R^2(L^2+\rho^2)}\mathcal{T}
\eeq
It turns out that the matrix $\left(g^{(0)}-b^{(0)}\right)^{-1}$ in (\ref{Yexp2}) can be written as
\beq
\left(g^{(0)}-b^{(0)}\right)^{-1}=\mathcal{C}^{-1}+\mathcal{J} = 
\mathcal{G}^{-1}+\frac{L^2}{R^2(L^2+\rho^2)}\mathcal{T} + \mathcal{J}
\eeq
Since the whole dependence on the deformation parameters is encoded in $\mathcal{T}$ and
$\mathcal{J}$, the $\hat{\gamma}_i \to 0$ limit is easily understood. 

Now a long but straightforward calculation allows to determine the first order corrections 
$g^{(1)},b^{(1)},G^{(1)}$ as well as the second order ones $g^{(2)},b^{(2)},G^{(2)}$. 
Inserting in $\mathcal{L}_{DBI}$ we eventually find 

\bea\label{L2}
\mathcal{L}_{DBI}^{(0)}&=& -T_7 \rho^3 c_\theta s_\theta\non
\\
\mathcal{L}^{(1)}_{DBI} &=& T_7 \rho^3 c_\theta s_\theta \hat{\gamma}_1 F_{\phi_2 \phi_3}/R^2
\non \\
\mathcal{L}_{DBI}^{(2)}&=&
-T_7 \rho^3 c_\theta s_\theta\left[\frac{R^2}{2(L^2+\rho^2)}\,\mathcal{C}^{ab}\partial_a \chi\partial_b 
\chi+\frac{R^2}{2(L^2+\rho^2)}\,\mathcal{G}^{ab}\partial_a \varphi\partial_b \varphi \right.\non\\
&~&+\left.\frac{1}{4}F_{ab}F^{ab}+
\frac{L}{(L^2+\rho^2)}(\hat{\gamma}_2 F_{a\phi_3}-
\hat{\gamma}_3 F_{a\phi_2})\,\mathcal{G}^{ab}\partial_b\varphi\right]
\eea
where $F^{ab} \equiv \mathcal{C}^{ac}\mathcal{C}^{bd}F_{cd}$ and
$\mathcal{C}^{ac}$ is given in (\ref{Cinv}). 
The first order Lagrangian is a total derivative since our embedding $X_5=L$, $X_6=0$ is an exact  
solution of the equations of motion.

The Wess--Zumino Lagrangian starts with a second order term in $\varepsilon$ given by
\beq\label{lwz}
\mathcal{L}_{WZ}=T_7\,\frac{1}{2}P[C_4-C_2\wedge B]\wedge F\wedge F=
T_7\frac{(L^2+\rho^2)^2}{R^4}\,\epsilon^{ijk}\partial_\rho A_i\partial_j A_k
\eeq
where we use latin indices to indicate coordinates on the three--sphere parametrized by
$(\theta,\phi_2,\phi_3)$, $A_i$ is the flux potential on
it and $\epsilon^{ijk}$ is the Levi--Civita tensor density ($\epsilon^{\theta 23} =1$). 
This term turns out to be independent of the deformation parameters since the combination 
$(C_4-C_2\wedge B)$ at lowest order  
gives exactly the 4--form of the ${\rm AdS}_5\times {\rm S}^5$ undeformed geometry.

Determining the equations of motion from the previous Lagrangian is now an easy task.
Introducing the fixed vector 
\beq
v^a = \hat{\g}_2 \d^a_3 - \hat{\g}_3 \d^a_2
\label{vector}
\eeq
for the $\chi$ and $\varphi$ scalars we find
\beq\label{eomchi}
\partial_a\left[\sqrt{-det(\mathcal{G})}\,\left( \frac{R^2}{(L^2+\rho^2)}\,
\mathcal{G}^{ab}\,+\, \frac{L^2}{(L^2+\rho^2)^2} v^a v^b \right) 
\partial_b\chi \right]=0
\eeq
\beq\label{eomphi}
\partial_a\left[\sqrt{-det(\mathcal{G})}\,\frac{R^2}{(L^2+\rho^2)}\,
\mathcal{G}^{ab}\,\left( \partial_b\varphi + \frac{L}{R^2} v^c F_{bc} \right)\right] =0
\eeq
whereas, using (\ref{eomphi}) the equations of motion for the gauge fields take the form
\bea\label{eoma}
&&\partial_a\left[\sqrt{-det(\mathcal{G})}\,\mathcal{G}^{ac}\mathcal{G}^{bd}\,F_{cd}\right]-
\frac{4\rho(L^2+\rho^2)}{R^4}\epsilon^{bjk}\partial_j A_k \\
&&~~~~~~~~~-\sqrt{-det(\mathcal{G})}\,\frac{L}{(L^2+\rho^2)}\, v^d \partial_d \,
\left[ \mathcal{G}^{bc}\,\left( \partial_c \varphi + \frac{L}{R^2} v^f
F_{c f}\right) \right]=0 \non
\eea
It is interesting to note that the equations of motion depend only on the deformation
parameters $\hat{\gamma}_2$ and $\hat{\gamma}_3$ hidden in the vector $v$. In fact, at this order 
the dependence on the parameter $\hat{\gamma}_1$ associated to the torus inside the D7 worldvolume 
completely cancels between the factors $\sqrt{-det(g - b + \varepsilon F)}$ and $1/\sqrt{G}$.
 
The scalar fluctuation $\chi$ along the direction where the 
branes are located at distance $L$ decouples from the rest. The scalar $\varphi$, instead,
interacts non--trivially with the worldvolume gauge fields through terms proportional to
the deformation parameters.

The vector $v$ has non--vanishing components only on the three--sphere and selects there
a fixed direction. As a consequence, the equations of motion (\ref{eomchi} -- \ref{eoma}) loose
$SO(4)$ invariance. 

\vskip 15pt
As a first application we consider the $L=0$ conformal case. The vibration of the brane 
is given by $X_5=\varepsilon \,\chi(\sigma^a)$ and $X_6=\varepsilon \,\varphi(\sigma^a)$.
The equations of motion reduce to
\bea
&&\partial_a\left[\sqrt{-det(\mathcal{G})}\,\frac{R^2}{\rho^2}\,\mathcal{G}^{ab}\,
\partial_b\Psi\right]=0\non\\
&&\partial_a\left[\sqrt{-det(\mathcal{G})}\,\mathcal{G}^{ac}\mathcal{G}^{bd}\,F_{cd}\right]
-\frac{4\rho^3}{R^4}\epsilon^{bjk}\partial_j A_k=0\,.
\eea
where $\Psi \equiv (\varphi,\chi)$ and $\mathcal{G}^{ab}$ is the inverse of the matrix 
(\ref{auxmet}) evaluated at $L=0$.
We see that the dependence on the deformation parameters disappears completely 
and the equations of motion reduce to the ones of the undeformed case \cite{KMMW}.
In particular, the scalar and gauge fluctuations decouple. Written explicitly, 
the scalar equations read 
\bea\label{undeformedeq}
\frac{R^4}{\rho^4}\partial^\mu\partial_\mu \Psi+\frac{1}{\rho^3}\partial_\rho(\rho^3\partial_\rho \Psi)+
\frac{1}{\rho^2}\Delta_{{\rm S}^3}\Psi=0
\eea
where 
\beq
\Delta_{{\rm S}^3} \Psi \equiv \frac{1}{c_\theta s_\theta} \pa_\theta (c_\theta s_\theta \pa_\theta \Psi)
+ \frac{1}{c_\theta^2} \pa_2^2 \Psi + \frac{1}{s_\theta^2} \pa_3^2 \Psi
\eeq
is the Laplacian on the unit 3--sphere 
($\pa_2 \equiv \pa_{\phi_2}$, $\pa_3 \equiv \pa_{\phi_3}$).   

According to the results in \cite{KK,KMMW} the corresponding ${\rm AdS}_5$ masses  
are above the Breitenlohner--Freedman bound \cite{BF}.
This is a further check of the stability of our brane configuration.

\section{The mesonic spectrum}

We now concentrate on the more general situation
$X_5=L+\varepsilon \,\chi(\sigma^a)$, $X_6=\varepsilon \,\varphi(\sigma^a)$ and
solve the equations of motion (\ref{eomchi} -- \ref{eoma}) 
for scalar and vector modes. We write the abelian flux in terms of its potential one--form,
$F_{ab} = \pa_a A_b - \pa_b A_a$, and choose
the Lorentz gauge $\pa_\mu A^\mu=0$ on the spacetime components.

We find convenient to introduce covariant derivatives on the unit three--sphere $(\theta, \phi_2,
\phi_3)$. Given its metric $g = diag(1, c_\theta^2, s_\theta^2)$, we have $\nabla_i V^j = 
\pa_i V^j + \Gamma_{ik}^j V^k$ with the only non--vanishing components being 
$\Gamma_{22}^\theta = - \Gamma_{33}^\theta = c_\theta s_\theta$, $\Gamma_{2\theta}^2 = 
-\frac{s_\theta}{c_\theta}$ and $\Gamma_{3\theta}^3 = \frac{c_\theta}{s_\theta}$.   
 
In order to simplify the equations we introduce the special operators 
\bea\label{ogamma}
&& \mathcal{O}_{\hg} \equiv \frac{R^4}{(L^2+\rho^2)^2}\partial^\nu\partial_\nu+
\frac{1}{\rho^3}\partial_\rho(\rho^3\partial_\rho)+
\frac{1}{\rho^2}\, \frac{1}{\sqrt{g}} \pa_i (\sqrt{g}\pa^i) +\frac{L^2}{(L^2+
\rho^2)^2}(\hat{\gamma}_2\partial_3-\hat{\gamma}_3\partial_2)^2
\non\\
&& \tilde{\mathcal{O}}_{\hg} \equiv \frac{R^4}{(L^2+\rho^2)^2}\,\partial^{\nu}\partial_\nu+
\frac{1}{\rho 
(L^2+\rho^2)^2}\partial_{\rho}\left[\rho (L^2+\rho^2)^2 \partial_\rho\right]
+\frac{1}{\rho^2}\nabla_l \nabla^l 
\non\\
&~&~~~~~~+\frac{L^2}{(L^2+\rho^2)^2}(\hat{\gamma}_2\partial_3-\hat{\gamma}_3\partial_2)^2
\eea
along with their undeformed versions $\mathcal{O}_0 \equiv \mathcal{O}_{\hg}|_{\hat{\gamma}_2 = 
\hat{\gamma}_3 =0}$, $\tilde{\mathcal{O}}_0 \equiv \tilde{\mathcal{O}}_{\hg}|_{\hat{\gamma}_2 = 
\hat{\gamma}_3 =0}$.
 
Equation (\ref{eomchi}) for the $\chi$ mode then takes the compact form 
\beq\label{seom}
\mathcal{O}_{\hg}\,\chi=0
\eeq
whereas equation (\ref{eomphi}) can be rewritten as
\beq\label{1eq}
\mathcal{O}_0\,\Phi-
\frac{L}{R^2}(\hat{\gamma}_2\partial_3-\hat{\gamma}_3\partial_2)
\left[\frac{1}{\rho^3}\partial_\rho(\rho^3 A_\rho)+\frac{1}{\rho^2}\nabla_l A^l \right]=0
\eeq
where we have defined
\beq\label{Phi}
\Phi\equiv \varphi + \frac{L}{R^2} v^a A_a = 
\varphi+\frac{L}{R^2}(\hat{\gamma}_2 A_3-\hat{\gamma}_3 A_2) 
\eeq

Equations (\ref{eoma}) for the vector modes come into three classes,
according to $b$ being in Minkowski, or $b = \rho$ or 
$b=i \equiv \{\theta,\,\phi_2,\,\phi_3\}$. We list the three cases.

\begin{itemize}
\item $b$ in Minkowski: For $b=\mu$ and expressing the $F$ flux in terms of its one--form
potential, equation (\ref{eoma}) becomes
\beq
\mathcal{O}_{\hg}\,A_\mu-\partial_\mu\left[\frac{1}{\rho^3}\partial_\rho(\rho^3 A_\rho)+
\frac{1}{\rho^2}\nabla_l A^l+
\frac{L R^2}{(L^2+\rho^2)^2}(\hat{\gamma}_2\partial_3-\hat{\gamma}_3\partial_2)\,\Phi\right]=0
\eeq
with $\Phi$ defined in (\ref{Phi}).

We apply $\pa^\mu$ to this equation and sum over $\mu$. Using $[\pa^\mu, \mathcal{O}_{\hg}]=0$
and Lorentz gauge, solutions corresponding to non--trivial dispersion relations ($k^2 \neq 0$) 
satisfy
\beq\label{2eq}
\left[\frac{1}{\rho^3}\partial_\rho(\rho^3 A_\rho)+\frac{1}{\rho^2}\nabla_l A^l+
\frac{L R^2}{(L^2+\rho^2)^2}(\hat{\gamma}_2\partial_3-\hat{\gamma}_3\partial_2)\Phi \right]=0
\quad , \quad 
\mathcal{O}_{\hg}\,A_\mu=0
\eeq

\item $b=\rho$: Again, expressing the flux in terms of the vector potential we obtain
\beq
\mathcal{O}_{\hg}\,A_\rho-
\left[\frac{1}{\rho^3}\partial_\rho(\rho^3 \partial_\rho A_\rho)+\frac{1}{\rho^2}\partial_\rho 
\nabla_l A^l+
\frac{L R^2}{(L^2+\rho^2)^2}(\hat{\gamma}_2\partial_3-\hat{\gamma}_3\partial_2)
\partial_\rho \Phi\right]=0
\eeq

\item $b=i$: On the internal $\tilde{{\rm S}}^3$ sphere we have 
\bea\label{rho}
&&\tilde{\mathcal{O}}_{\hg}\,A_j-
\frac{1}{\rho^2}\left(\nabla_l \nabla_j A^l+\frac{4\rho^2}{L^2+\rho^2}
\frac{1}{c_\theta s_\theta} \epsilon_{jlm}\nabla^l A^m\right)
\\
&~&~~-\frac{1}{\rho (L^2+\rho^2)^2}\partial_\rho\left[\rho (L^2+\rho^2)^2 \partial_j A_\rho\right]
-\frac{L R^2}{(L^2+\rho^2)^2}(\hat{\gamma}_2\partial_3-\hat{\gamma}_3\partial_2) \,
\partial_j\Phi =0 \non
\eea
where we have used $\frac{1}{\sqrt{g}} \pa_i (\sqrt{g} F^{ij}) = \nabla_i F^{ij}= 
\nabla_i \nabla^i A^j - \nabla_i \nabla^j A^i$. 
\end{itemize}

\noindent
Now, collecting all the equations and using the first of (\ref{2eq}) 
in (\ref{1eq}) the system of coupled equations we need solve is
\bea\label{sisfin}
(0)\, &&\mathcal{O}_{\hg} \chi =0  \qquad  ; \qquad \mathcal{O}_{\hg}\,A_\mu=0\non\\
\\
(1)\,&&\mathcal{O}_{\hg}\,\Phi=0 \non\\ 
\non\\
(2)\, && \left[\frac{1}{\rho^3}\partial_\rho(\rho^3 A_\rho)+
\frac{1}{\rho^2}\nabla^l A_l+
\frac{L R^2}{(L^2+\rho^2)^2}(\hat{\gamma}_2\partial_3-\hat{\gamma}_3\partial_2)\,
\Phi\right]=0\non\\
\non\\
(3)\,&&\mathcal{O}_{\hg}\,A_\rho-
\left[\frac{1}{\rho^3}\partial_\rho(\rho^3 \partial_\rho A_\rho)+\frac{1}{\rho^2}\partial_\rho 
\nabla^l A_l+
\frac{L R^2}{(L^2+\rho^2)^2}(\hat{\gamma}_2\partial_3-\hat{\gamma}_3\partial_2)
\partial_\rho \Phi\right]=0\non\\
\non\\
(4)\,&&\tilde{\mathcal{O}}_{\hg}\,A_j-
\frac{1}{\rho^2}\left(\nabla_l \nabla_j A^l+\frac{4\rho^2}{L^2+\rho^2} \frac{1}{c_\theta s_\theta}
\epsilon_{jlm}\nabla^l A^m\right)
\non\\
&~&~-\frac{1}{\rho (L^2+\rho^2)^2}\partial_\rho\left[\rho (L^2+\rho^2)^2 \partial_j A_\rho\right]
-\frac{L R^2}{(L^2+\rho^2)^2}(\hat{\gamma}_2\partial_3-\hat{\gamma}_3\partial_2)
\partial_j\Phi=0\non
\eea
Equations $(1)$--$(4)$ exhibit a non--trivial 
interaction between the scalar $\Phi$ and the components of the vector potential along the
internal directions. The modes $\chi$ and $A_\mu$ instead decouple.

It is convenient to search for solutions expanded in spherical harmonics on ${\rm S}^3$. 
Scalar spherical harmonics are a complete set of functions $\mathcal{Y}_l^{m_2,m_3}$ 
in the $\left(\frac{l}{2},\frac{l}{2}\right)$ representation of $SO(4)$ and  
with definite $U(1)\times U(1)$ quantum numbers $(m_2,m_3)$
satisfying $|m_2+m_3| = |m_2-m_3| =l - 2k$, $l,k =0,1,\dots$. For fixed $l$ the degeneracy
is $(l+1)^2$. Their defining equations are
\footnote{For their explicit realization see for instance \cite{leq,hsz}.}
\bea\label{Yprop}
\Delta_{{\rm S}^3}\,\mathcal{Y}_l^{m_2,m_3}&=&
-l(l+2) \, \mathcal{Y}_l^{m_2,m_3}\non\\
\frac{\partial}{\partial \phi_{2,3}}\,\mathcal{Y}_l^{m_2,m_3}&=&
i m_{2,3} \, \mathcal{Y}_l^{m_2,m_3}
\eea
Vector spherical harmonics come into three classes. Choosing them to be also eigenfunctions of
$\frac{\partial}{\partial \phi_{2,3}}$ we have longitudinal harmonics
$\mathcal{H}_i=\nabla_i \mathcal{Y}_l^{m_2,m_3}$, $l \geq 1$  which 
are in the $(\frac{l}{2}, \frac{l}{2})$ 
representation of $SO(4)$ with $(m_2,m_3)$ ranging as before. 
Transverse harmonics are $\mathcal{M}^{+}_i \equiv \mathcal{Y}_i^{(l,m_2,m_3);+}$ with $l\geq 1$ 
in the $\left(\frac{l- 1}{2},\frac{l+1}{2}\right)$ and 
$\mathcal{M}^{-}_i \equiv \mathcal{Y}_i^{(l,m_2,m_3);-}$ with $l\geq 1$ 
in the $\left(\frac{l+ 1}{2},\frac{l-1}{2}\right)$. Their degeneracy is $l(l+2)$ and it is
counted by $|m_2+m_3| = l \pm 1-2k, |m_2-m_3| =l \mp 1- 2k$.  
These harmonics satisfy
\bea\label{proM}
\nabla_i\nabla^i \mathcal{M}^\pm_j-R^k_j \mathcal{M}^\pm_k&=&-(l+1)^2 \, \mathcal{M}^\pm_j\non\\
\epsilon_{ijk}\nabla^j \mathcal{M}^{\pm;k}&=& \pm \,\sqrt{g} \,(l+1)\, \mathcal{M}^\pm_i\non\\
\nabla^i\mathcal{M}^\pm_i&=&0\non\\
\frac{\partial}{\partial \phi_{2,3}}\,\mathcal{M}_i^{\pm}&=&
i m_{2,3} \, \mathcal{M}_i^{\pm}
\eea
where $\sqrt{g} = c_\th s_\th$ is the square root of the determinant of the metric on 
${\rm S}^3$, whereas $R^i_j=2\delta^i_j$ is the Ricci tensor.

As in the undeformed case \cite{KMMW} we require the solutions to be regular at the
origin ($\rho=0$), normalizable and small enough to justify the quadratic approximation. 
All these conditions are used to select the actual mass spectrum of the mesonic excitations.

\subsection{The decoupled modes}

\subsubsection{The scalar mode $\chi$}

We start solving the equation for the decoupled scalar $\chi$. Using the general identity 
$\frac{1}{\sqrt{g}} \pa_i (\sqrt{g}\pa^i s) = \nabla_i \nabla^i s$ 
valid for any scalar $s$, the equation ${\cal O}_{\hat{\g}} \chi =0$ reads explicitly 
\beq
\label{chieq}
\frac{R^4}{(L^2+\rho^2)^2}\partial^\nu\partial_\nu \chi +\frac{1}{\rho^3}
\partial_\rho(\rho^3\partial_\rho \chi)+
\frac{1}{\rho^2}\nabla_l \nabla^l \chi +\frac{L^2}{(L^2+\rho^2)^2}(\hat{\gamma}_2\partial_3-
\hat{\gamma}_3\partial_2)^2 \chi=0
\eeq
We look for single--mode solutions of the form
\beq
\chi(\sigma^a)=r(\rho)\, e^{i k x}\,\mathcal{Y}_l^{m_2,m_3}(\theta,\phi_2,\phi_3)
\eeq
Inserting in (\ref{chieq}) we obtain an equation for $r(\rho)$ that, 
after the redefinitions
\beq\label{red}
\varrho=\frac{\rho}{L}\,,\qquad\qquad \hat{\Gamma}^2=
-\frac{k^2 R^4}{L^2}-(\hat{\gamma}_2 m_3-\hat{\gamma}_3 m_2)^2=
\bar{M}^2-(\hat{\gamma}_2 m_3-\hat{\gamma}_3 m_2)^2\,,
\eeq
becomes
\beq
\partial_\varrho^2 r+\frac{3}{\varrho}\partial_\varrho r+
\left[\frac{\hat{\Gamma}^2}{(1+\varrho^2)^2}-\frac{l(l+2)}{\varrho^2}\right]r=0
\eeq
This has exactly the same structure of the equation found in the undeformed case \cite{KMMW}. 
The only difference is the presence of the deformation parameters in $\hat{\Gamma}^2$
which in the undeformed case reduces simply to $\bar{M}^2$.
Following what has been done in that case \cite{KMMW} we find that the general solution is
\beq
r(\rho) = \rho^l (L^2 + \rho^2)^{-\alpha} F(-\alpha, -\alpha +l+1; l+2; - \rho^2/L^2)
\eeq
where $F$ is the hypergeometric function and $\alpha=\frac{-1+ \sqrt{1+\hat{\Gamma}^2}}{2}$. 
This solution satisfies the conditions of regularity and normalizability 
if the quantization condition 
\beq\label{qc}
\hat{\Gamma}^2=4(n+l+1)(n+l+2)\qquad \qquad n \in N \, , \quad n,l \geq 0
\eeq
is imposed.
Using (\ref{red}) and $M^2=-k^2$, the mass spectrum of scalar mesons then follows
\beq\label{chispectrum}
M_\chi(n,l,m_2,m_3)=\frac{2 L}{R^2}\sqrt{(n+l+1)(n+l+2)+
\left(\frac{\hat{\gamma}_2 m_3-\hat{\gamma}_3 m_2}{2}\right)^2}
\eeq
with $n,l \geq 0$ and $|m_2+m_3|=|m_2-m_3|=l-2k$, $k$ a non--negative integer.

We see that the deformation parameters induce a non--trivial dependence of the mass spectrum 
on the two $U(1)$ quantum numbers $(m_2,m_3)$, so breaking the degeneracy of the undeformed case.

The mass spectrum is smoothly related to the one of the undeformed case for $\hat{\gamma}_i
\to 0$. 

\subsubsection{The Type II modes}

We look for excitations of the form 
\beq
A_\mu(\sigma^a)=\zeta_\mu\,Z_{II}(\rho)\,e^{i k x}\,
\mathcal{Y}^{m_2,m_3}_l(\theta,\phi_2,\phi_3) \quad , \quad k \cdot \zeta=0
\eeq
Following the classification introduced in \cite{KMMW} for the undeformed case
we call them Type II modes.
The equation  $\mathcal{O}_{\hg} A_\mu=0$ in (\ref{sisfin}) yields to
\beq
\frac{R^4}{(L^2+\rho^2)^2}\partial^\nu\partial_\nu A_\mu+\frac{1}{\rho^3}
\partial_\rho(\rho^3\partial_\rho A_\mu)+
\frac{1}{\rho^2}\nabla_l \nabla^l A_\mu+\frac{L^2}{(L^2+\rho^2)^2}(\hat{\gamma}_2\partial_3-
\hat{\gamma}_3\partial_2)^2 A_\mu=0
\eeq
This is exactly the same equation as the one for the scalar mode $\chi$.
Therefore, for each component $A_\mu$ we follow the same strategy of subsection 5.1.1
and find the mass spectrum 
\beq
M_{II}(n,l,m_2,m_3)=\frac{2 L}{R^2}\sqrt{(n+l+1)(n+l+2)+\left(\frac{\hat{\gamma}_2 m_3-
\hat{\gamma}_3 m_2}{2}\right)^2} 
\eeq
with $n,l \geq 0$ and $|m_2+m_3|=|m_2-m_3|=l-2k$.

Even for this type of vector fluctuations the spectrum is smoothly related to the undeformed
one for $\hat{\gamma}_{i} \to 0$.

\vspace{3.3mm}

\subsection{The coupled modes}

Having performed the field redefinition (\ref{Phi}) we solve the coupled equations $(1)$--$(4)$ 
by considering elementary fluctuations of $\Phi$, $A_\rho$ and $A_i$.

\subsubsection{The Type I modes}

Being in a different representation the harmonics $\mathcal{M}^\pm_i$ do not mix with the others. 
Therefore we can make the ansatz \footnote{We note that if we were to follow closely the 
classification of \cite{KMMW} we would call 
Type I modes the elementary modes with $\varphi=0$, i.e. with no fluctuations along the
$X^6$ coordinate. However, given the structure of the equations of motion, in our case we find
the definition (\ref{Imod}) more convenient. In any case, the two definitions coincide for 
$\hat{\g}_i=0$.}
\beq\label{Imod}
\Phi=0,\qquad A_\rho=0,\qquad A_i(\sigma^a)=
Z^\pm_{I}(\rho)\,e^{i k x}\,\mathcal{M}^\pm_i(\theta, \phi_2,\phi_3)
\eeq
By using the identity $\nabla_i A^i=0$ as follows from (\ref{proM}), equations $(1)$, $(2)$ and $(3)$ 
in (\ref{sisfin}) are identically satisfied whereas eq. $(4)$ reads
\beq
\tilde{\mathcal{O}}_{\hg}\,A_j-
\frac{1}{\rho^2}\left(\nabla_l \nabla_j A^l+\frac{4\rho^2}{L^2+\rho^2} \frac{1}{c_\theta s_\theta}
\epsilon_{jlm}\nabla^l A^m\right) =0
\eeq
Considering the explicit expression for the operator $\tilde{\cal O}_{\hat{\g}}$ 
in (\ref{ogamma}) and using
properties (\ref{proM}) we find that $Z^\pm_{I}(\rho)$ is a solution of the equation
\beq
\frac{1}{\varrho} \pa_\varrho \left[ \varrho (\varrho^2 + 1)^2 \pa_\varrho Z^\pm_{I} \right]
+ \left[ \hat{\G}^2 - \frac{(\varrho^2 + 1)^2}{\varrho^2} (l+1)^2 \mp 4(\varrho^2 + 1)(l+1)
\right] \, Z^\pm_{I} =0
\eeq  
where we have used the definitions (\ref{red}). This is formally the same equation as the one
of the undeformed case, except for the different definition of $\hat{\G}^2$. Therefore, following
the same steps \cite{KMMW} we find that the solutions are still hypergeometric functions 
\bea
&& Z^+_I(\rho) = \rho^{l+1} (\rho^2 + L^2)^{-\a -1} F(l+2 -\a, -1-\a;l+2; -\rho^2/L^2)  \non \\
&& Z^-_I(\rho) = \rho^{l+1} (\rho^2 + L^2)^{-\a -1} F(l-\a, 1-\a;l+2; -\rho^2/L^2)
\eea
where $\a = \frac{-1 + \sqrt{1 +\hat{\G}^2}}{2}$.
Requiring them to be regular at infinity we obtain the following quantization conditions 
\bea
&& \hat{\G}^2_+ = 4(n+l+2)(n+l+3) \non \\
&& \hat{\G}^2_- = 4(n+l)(n+l+1) \qquad \qquad n \geq 0
\eea 
As a consequence the mass spectrum reads
\bea
M_{I,+} = \frac{2L}{R^2} \sqrt{ (n+l+2)(n+l+3) + \left( \frac{\hat{\g}_2 m_3 - \hat{\g}_3 m_2}{2}
\right)^2} \quad && 
\left\{
\begin{array}{l}
    |m_2+m_3|=l - 1 -2k \\
    |m_2-m_3|=l +1 -2k 
\end{array}
\right.
\non \\
~~~\non \\
M_{I,-} = \frac{2L}{R^2} \sqrt{ (n+l)(n+l+1) + \left( \frac{\hat{\g}_2 m_3 - \hat{\g}_3 m_2}{2}
\right)^2} \qquad \quad && 
\left\{
\begin{array}{l}
  |m_2+m_3|=l + 1 -2k \\
  |m_2-m_3|=l - 1 -2k 
\end{array}
\right.\non\\
\eea
where $l \geq 1$ and $k$ is a non--negative integer.

\subsubsection{The Type III modes}
\vspace{1.2mm}

Finally, we consider the following fluctuations
\bea\label{modesIII}
\Phi(\sigma^a)&=& X_{III}(\rho)\, e^{i k x}\,\mathcal{Y}^{m_2,m_3}_l(\theta,\phi_2,\phi_3) 
\non\\
A_\rho(\sigma^a)&=&Y_{III}(\rho)\, e^{i k x}\,\mathcal{Y}^{m_2,m_3}_l(\theta,\phi_2,\phi_3) \quad 
 \\
A_i(\sigma^a)&=&Z_{III}(\rho)\, e^{i k x}\,\nabla_i\mathcal{Y}^{m_2,m_3}_l(\theta,\phi_2,\phi_3) \equiv
\nabla_i A(\sigma^a) \non
\eea
with $l \geq 1$. We note that $l=0$ corresponds to having $A_i=0$. 
We will comment on this particular case at the end of this Section. 
 
Inserting in (\ref{sisfin}) and using the identities (\ref{Yprop}) for the scalar harmonics,
after a bit of algebra the equations $(1)$--$(4)$ can be rewritten as 
\bea\label{IIIm}
(1)\,&&\left[\frac{R^4}{(L^2+\rho^2)^2} \pa^\nu \pa_\nu + \frac{1}{\rho^3} \pa_\rho\left(
\rho^3 \pa_\rho \right) -
\frac{l(l+2)}{\rho^2} - \frac{L^2}{(L^2+\rho^2)^2}(\hat{\gamma}_2m_3-
\hat{\gamma}_3m_2)^2\right]\,\Phi = 0
\non\\
(2)\,&&\frac{1}{\rho^3}\partial_\rho(\rho^3 A_\rho)-\frac{l(l+2)}{\rho^2}A 
+i \frac{L R^2}{(L^2+\rho^2)^2}(\hat{\gamma}_2 m_3-\hat{\gamma}_3 m_2) \, \Phi=0 \non \\
\non\\
(3)\,&&\frac{R^4}{(L^2+\rho^2)^2} \pa^\nu \pa_\nu A_\rho +
\frac{1}{\rho^2} \pa_\rho \left( \frac{1}{\rho} \pa_\rho(\rho^3 A_\rho)\right)
\non \\
&~&~~~~~~~~~~- \left[ \frac{l(l+2)}{\rho^2} +
\frac{L^2}{(L^2 + \rho^2)^2}(\hat{\gamma}_2 m_3 - \hat{\gamma}_3 m_2)^2 \right] \,A_\rho
\non \\
&~&~~~~~~~~~~~~~~~~~~~~~~~ 
+ 2i LR^2\frac{(L^2-\rho^2)}{\rho(L^2+\rho^2)^3}( \hat{\gamma}_2m_3-\hat{\gamma}_3m_2)\, \Phi =0
\non\\
\non\\
(4)\,&&\frac{R^4}{(L^2+\rho^2)^2} \pa^\nu \pa_\nu A+
\frac{1}{\rho (L^2+\rho^2)^2} \pa_\rho\left(\rho (L^2+\rho^2)^2 \pa_\rho A\right)
\non \\
&&~~~~~- \frac{L^2}{(L^2+\rho^2)^2} (\hat{\gamma}_2m_3-\hat{\gamma}_3m_2)^2 \, A
- \frac{1}{\rho (L^2+\rho^2)^2}
\partial_\rho\left[\rho (L^2+\rho^2)^2  A_\rho\right] 
\non\\
&&~~~~~~~~~~~~~~~~~~~~
-i\frac{L R^2}{(L^2+\rho^2)^2}
(\hat{\gamma}_2m_3-\hat{\gamma}_3m_2)\, \Phi =0  
\eea
It is worth mentioning that eq. $(1)$ in (\ref{sisfin}) contains the operator 
$\frac{1}{\sqrt{g}} \pa_i ( \sqrt{g} \pa^i )$ which acts differently on 
scalars and spherical vectors. Therefore, when this operator is applied on $\Phi =\varphi + 
\frac{L}{R^2}(\hg_2 A_3 - \hg_3 A_2)$, in principle one 
should split it as acting on $\varphi$ and $A_i$ separately. However, since in the present
case $A_i = \nabla_i A$, exploiting the algebra of covariant derivatives and the properties of
scalar harmonics in (\ref{modesIII}), it is easy to show that 
\beq
\frac{1}{\sqrt{g}} \pa_i ( \sqrt{g} \pa^i \nabla_j A ) = \nabla_i \nabla^i \nabla_j A
- 2 \nabla_j A = - l(l+2) \nabla_j A
\eeq
This is exactly the same relation satisfied by the scalar $\varphi$, so we are led to 
$\frac{1}{\sqrt{g}} \pa_i ( \sqrt{g} \pa^i \Phi) = -l(l+2) \Phi$. This confirms 
that considering $\Phi$ as an elementary scalar fluctuation is a consistent procedure.  
 
Equations (\ref{IIIm}) are four equations for three unknowns 
$X_{III}, Y_{III}, Z_{III}$ and lead to non--trivial
solutions only if they are compatible. Indeed it turns out that equation $(4)$ is identically
satisfied once the others are. We then concentrate on the first three equations.

We first solve equation $(1)$. By observing that it is identical to the equation for the scalar 
$\chi$ (see eq. (\ref{chieq})) we immediately obtain 
\beq
X_{III}(\rho) =  \rho^l (L^2 + \rho^2)^{-n-l-1} F(-(n+l+1), -n; l+2; - \rho^2/L^2)
\label{XIII}
\eeq
where the quantization condition (\ref{qc}) has been used. As a consequence, the mass spectrum is 
\beq
M_\Phi(n,l,m_2,m_3)=\frac{2 L}{R^2}\sqrt{(n+l+1)(n+l+2)+
\left(\frac{\hat{\gamma}_2 m_3-\hat{\gamma}_3 m_2}{2}\right)^2}
\label{MPhi}
\eeq
where $n \geq 0$, $l \geq 1$ and $|m_2+m_3| = |m_2-m_3|= l-2k$. 

Equation $(2)$ can be used to express the mode $A$ in terms of $\Phi$
and $A_\rho$. Inserting the expressions (\ref{modesIII}) we obtain
\beq
Z_{III} = \frac{1}{l(l+2)} \left[ \frac{1}{\rho} \pa_\rho(\rho^3 Y_{III}) + 
i \frac{LR^2 \rho^2}{(L^2+\rho^2)^2} (\hat{\gamma}_2 m_3-\hat{\gamma}_3 m_2) X_{III} \right]
\label{A}
\eeq
We then consider equation $(3)$ which exhibits an actual coupling between $X_{III}$ and $Y_{III}$. 
In order to solve for $Y_{III}$ given the solution (\ref{XIII}) for $X_{III}$ we set
\beq
Y_{III}(\varrho) = \varrho^{l-1}(1+\varrho^2)^{-\a} \, P(\varrho)
\eeq
Using the definitions (\ref{red}) together with the quantization
condition (\ref{qc}) and defining $y \equiv -\varrho^2$, after some algebra the equation for $P$ reads
\bea
&& y(1-y) P''(y) + \left[ (l+2) + (2n+l) \,y \right] P'(y) - n(n+l+1) P(y) 
\non\\
&~&~~~~~~~~~~~~~~= \eta \, \frac{(1+y)}{(1-y)^2} F(-(n+l+1), -n; l+2;y)
\label{inhomog}
\eea
where we have defined $\eta \equiv i\frac{R^2}{2L^2} (\hat{\g}_2m_3 - \hat{\g}_3m_2)$.
This is an inhomogeneous hypergeometric equation whose source is a polynomial of degree $n$,
solution of the corresponding homogeneous equation.
The most general solution is then of the form
\beq
P(y) = c \, F(-(n+l+1), -n; l+2;y) +  \bar{P}(y)
\eeq
for arbitrary constant $c$, 
where $\bar{P}$ is a particular solution of (\ref{inhomog}). 
Exploiting the general identity
\bea
&&(1-y) \, F'(-(n+l+2),-n;l+1;y) \, + \, (n+l+2) \, F(-(n+l+2),-n;l+1;y) 
\non \\
&&~~~~~~~~~~~~~= \frac{(n+l+1)(n+l+2)}{(l+1)}\, F(-(n+l+1),-n;l+2;y)
\eea
valid for hypergeometric functions with integer coefficients, 
it is easy to show that a solution is given by
\beq
\bar{P}(y) = \eta \, \frac{(l+1)}{(n+l+1)(n+l+2)} \, \frac{F(-(n+l+2),-n;l+1;y)}{1-y}
\eeq
The general solution of equation $(3)$ is then
\bea
Y_{III}(\rho) &=& \rho^{l-1}(L^2+\rho^2)^{-n-l-2} \Big[ c \,
(L^2+\rho^2)\, F(-(n+l+1), -n; l+2; -\rho^2/L^2) \non \\
&+& \eta \, \frac{(l+1)}{(n+l+1)(n+l+2)} F(-(n+l+2),-n;l+1;-\rho^2/L^2) \Big]
\label{YIII}
\eea
This solution is regular at the origin and not divergent for $\rho \to \infty$. 
Due to the quantization condition (\ref{qc}) the corresponding mass spectrum is still given by
\beq
M_{III}(n,l,m_2,m_3)=\frac{2 L}{R^2}\sqrt{(n+l+1)(n+l+2)+
\left(\frac{\hat{\gamma}_2 m_3-\hat{\gamma}_3 m_2}{2}\right)^2}
\eeq
with $ n \geq 0$, $l \geq 1$ and  $|m_2+m_3| = |m_2-m_3|= l-2k$.

Before closing this Section we comment on the particular $l=m_2=m_3=0$ mode. 
In (\ref{modesIII}) this 
corresponds to turn off $A_i = \nabla_i A$ since $A(\s^a)$ is independent of the three--sphere
coordinates. Equation $(2)$ reduces to $\pa_\rho (\rho^3 A_\rho) =0$ which, together with the
condition of regularity at $\rho=0$, sets $A_\rho=0$. Equations $(3)$ and $(4)$ in (\ref{IIIm}) 
are then automatically satisfied, whereas eq. $(1)$ provides a non--trivial solution  
for $\Phi$ as given in (\ref{XIII}) with mass (\ref{MPhi}) where we set $l=m_2=m_3=0$ .

As a slightly different attitude we can consider the configuration with all the vector modes 
turned off ($Y_{III}= Z_{III}=0$) and study only scalar $\Phi$ fluctuations of the form 
(\ref{modesIII}). 
In this case $\Phi$ is still solution of equation $(1)$ but, as follows from the rest of
equations, it is constrained by the further condition
\beq
(\hat{\g}_2 m_3 - \hat{\g}_3 m_2) \Phi =0
\eeq
In general, for non--vanishing and distinct deformation parameters, non--trivial solutions 
can be found only for $m_2=m_3=0$, i.e. only the $U(1) \times U(1)$ zero--mode sector is
selected and the fluctuations are independent of $(\phi_2,\phi_3)$. A greater number of 
solutions, corresponding to the modes $m_2=m_3$, is instead allowed when $\hat{\g}_2=
\hat{\g}_3$, therefore in particular for the supersymmetric deformation. In any case, the
mass spectrum is given by
\beq
M_\Phi(n,l)=\frac{2 L}{R^2}\sqrt{(n+l+1)(n+l+2)}
\qquad n \ge 0 \quad l ~{\rm (even)} \geq 0
\eeq
and coincides with the undeformed mass.

\section{Analysis of the spectrum}

From the previous discussion it follows that the bosonic modes arising from the 
compactification of the D7--brane on the deformed $\tilde{{\rm S}}^3$ give rise to a  
mesonic spectrum which is given by 
\begin{itemize}
\item 2 scalars and 1 vector in the $(\frac{l}{2},\frac{l}{2})$ with $l\geq 0$, $|m_2 \pm m_3|=l-2k$ and mass
\bea
M_{\chi,\Phi,II}(n,l,m_2,m_3)&=&\frac{2 L}{R^2}\sqrt{(n+l+1)(n+l+2)+\left(\frac{\hat{\gamma}_2 m_3-
\hat{\gamma}_3 m_2}{2}\right)^2} \non
\eea
\item 1 scalar in the $(\frac{l}{2},\frac{l}{2})$ with $l\geq 1$, $|m_2 \pm m_3|=l-2k$ and mass
\bea
M_{III}(n,l,m_2,m_3)&=&\frac{2 L}{R^2}\sqrt{(n+l+1)(n+l+2)+\left(\frac{\hat{\gamma}_2 m_3-
\hat{\gamma}_3 m_2}{2}\right)^2} \non
\eea
\item 1 scalar in the $(\frac{l-1}{2},\frac{l+1}{2})$ with $l\geq 1$, $|m_2 \pm m_3|=l\mp 1-2k$ and mass
\bea
M_{I,+}(n,l,m_2,m_3)&=&\frac{2 L}{R^2}\sqrt{(n+l+2)(n+l+3)+\left(\frac{\hat{\gamma}_2 m_3-
\hat{\gamma}_3 m_2}{2}\right)^2} \non
\eea
\item 1 scalar in the $(\frac{l+1}{2},\frac{l-1}{2})$ with $l\geq 1$, $|m_2 \pm m_3|=l\pm 1-2k$ and mass
\bea
M_{I,-}(n,l,m_2,m_3)&=&\frac{2 L}{R^2}\sqrt{(n+l)(n+l+1)+\left(\frac{\hat{\gamma}_2 m_3-
\hat{\gamma}_3 m_2}{2}\right)^2} \non
\eea
\end{itemize}
for any $n \geq 0$. 
This matches exactly the bosonic content found in the undeformed case \cite{KMMW}. 
However, in this case the $\g$--deformation breaks $SO(4) \to U(1) \times U(1)$ and  
induces an explicit dependence of the mass spectrum on the
the quantum numbers $(m_2,m_3)$ with a pattern similar to the Zeeman effect
for atomic electrons where the constant magnetic field which breaks
$SU(2)$ rotational invariance down to $U(1)$ induces a dependence of the energy
levels on the azimuthal quantum number $m$ \footnote{A similar effect has been observed
in the case of backgrounds with $B$ fields turned on in Minkowski \cite{other3,nuovi}.}.  
 
The dependence on the deformation parameters disappears completely 
in the $m_2=m_3=0$ sector (or for $\hg_2=\hg_3$ and $m_2=m_3$) and the mass eigenvalues 
coincide with the ones of the undeformed theory. 
When $\hat{\g}_2 = \hat{\g}_3$ the mass spectrum acquires an
extra symmetry under the exchange of the two $U(1)$'s and an extra degeneracy corresponding
to $m_2 \to m_2 + m$, $m_3 \to m_3 +m$, $m$ integer. 

For any value of $\hg_i$ there are no tachyonic modes, so confirming the stability 
of our configuration. Moreover, massless states are absent and the spectrum has a 
mass gap given by
\beq
M_{gap}=2\sqrt{2}\frac{L}{R^2}
\eeq
This is exactly the mass gap present in the undeformed theory \cite{KMMW}.

In order to analyze in detail the mass splitting induced by the deformation 
and study how the modes organize themselves among the different eigenvalues
it is convenient to rewrite the mass of a generic eigenstate $X$ as 
\beq
M_X(n,l,m_2,m_3)= \sqrt{\left(M_X^{(0)}(n,l)\right)^2 +\frac{4 L^2}{R^4} \left(\Delta M(m_2,m_3)\right)^2}
\label{MX}
\eeq 
where $M_X^{(0)}$ is the undeformed mass, whereas
\beq\label{spl}
\Delta M(m_2,m_3)\equiv \left(\frac{\hat{\gamma}_2 m_3-
\hat{\gamma}_3 m_2}{2}\right)
\eeq
is the Zeeman--splitting term.

Since for any $l \geq 2$ the following mass degeneracy occurs
\beq
M^{(0)}_{\chi,\Phi,II}(n,l)=M^{(0)}_{III}(n,l)=M^{(0)}_{I,+}(n,l-1)=M^{(0)}_{I,-}(n,l+1)
\label{shift}
\eeq
for $\hat{\g}_i=0$ we have $8(l+1)^2$ bosonic degrees of freedom corresponding to the same mass 
eigenvalue.
For the particular values $l=0,1$ the number of states is reduced since for $l=0$ modes
$A_{(I,+)}$ and $A_{III}$ are both absent, whereas for $l=1$ $A_{(I,+)}$ is still absent. 
For any value of $l$ 
they match the bosonic content of massive ${\cal N}=2$ supermultiplets \cite{KMMW}. 
     
In the present case mass degeneracy occurs among states which satisfy the above condition and have 
the same value of $\Delta M(m_2,m_3)$. Therefore, having performed the $l$--shift for the
$(I,\pm )$ modes as in (\ref{shift}), we concentrate on the degeneracy in $\Delta M(m_2,m_3)$ 
for fixed values of $(n,l)$. 
It is convenient to discuss the $\hg_2=\hg_3$ and $\hg_2\neq\hg_3$ cases,
separately.

\vskip 12pt 
\noindent 
\underline{$\hg_2=\hg_3 \equiv\hg$}: This case includes the supersymmetric LM--theory. 
The deformation enters the mass spectrum only through the difference $(m_2-m_3)$
and the splitting term $\Delta M$ depends only on a single integer $j$
\bea\label{sumas}
&&l\quad\mbox{even}\qquad  2j\equiv |m_2-m_3|=0,2,\cdots,l  \qquad\quad\,\,\ \, \Delta M(j) =
\hg\,j\non\\
&&l\quad\mbox{odd}\qquad\,\,  2j+1\equiv |m_2-m_3|=1,3,\cdots,l  \qquad \Delta M(j) =
\hg\,\left(j+\frac{1}{2}\right)
\eea
Excluding for the moment the $l=0,1$ cases, for any given value of $2j$ and $2j+1$ the 
degeneracies of the corresponding mass levels are listed in Table
\ref{tabpar} and Table \ref{tabdis}, respectively. 

\begin{table}[h]
\begin{center}
\begin{tabular}{||*{3}{c|}|}
\hline
$\mbox{State}$ & $|m_2-m_3|=2j$ & $\mbox{Degeneracy}$ \\
\hline
\hline
$\chi,\,\Phi,\,A_{III}$ & 
\hspace{-3.5mm}
\begin{tabular}{c}
$0$\\
\hline
$\phantom{aaaa}2,4,\cdots,l\phantom{aaaa}$
\end{tabular}
\hspace{-3.5mm}
&
\hspace{-4mm}
\begin{tabular}{c}
$l+1$\\
\hline
$\phantom{aaaa}2(l+1)\phantom{aaaa}$
\end{tabular}
\hspace{-6mm}
$\phantom{.}$
\\
\hline
\hspace{-2mm}$A_\mu$ & 
\hspace{-3.5mm}
\begin{tabular}{c}
$0$\\
\hline
$\phantom{aaaa}2,4,\cdots,l\phantom{aaaa}$
\end{tabular}
\hspace{-3.5mm}
& \hspace{-4mm}
\begin{tabular}{c}
$l+1$\\
\hline
$\phantom{aaaa}2(l+1)\phantom{aaaa}$
\end{tabular}
\hspace{-6mm}
$\phantom{.}$
\\
\hline
$A_{I,+}$ & 
\hspace{-3.5mm}
\begin{tabular}{c}
$0$\\
\hline
$\phantom{aaaa}2,4,\cdots,l\phantom{aaaa}$
\end{tabular}
\hspace{-3.5mm}
&\hspace{-4mm}
\begin{tabular}{c}
$l-1$ \\
\hline
$\phantom{aaaa}2(l-1)\phantom{aaaa}$
\end{tabular}
\hspace{-6mm}
$\phantom{.}$
\\
\hline
$A_{I,-}$ & 
\hspace{-3.5mm}
\begin{tabular}{c}
$0$\\
\hline
$\phantom{aaaa}2,4,\cdots,l\phantom{aaaa}$
\end{tabular}
\hspace{-3.5mm}
&\hspace{-4mm}
\begin{tabular}{c}
$l+3$\\
\hline
$\phantom{aaaa}2(l+3)\phantom{aaaa}$
\end{tabular}
\hspace{-6mm}
$\phantom{.}$
\\
\hline
\end{tabular}
\end{center}
\caption{Degeneracy of states in the case $\hg_2=\hg_3$ and $l\geq 2$ even. 
The degeneracy in the third column refers to every single value of $j$.}
\label{tabpar}
\end{table}

\begin{table}[h]
\begin{center}
\begin{tabular}{||*{3}{c|}|}
\hline
$\mbox{State}$ & $|m_2-m_3|=2j+1$ & $\mbox{Degeneracy}$ \\
\hline
\hline
$\chi,\,\Phi,\,A_{III}$ & $1,3,\cdots,l$ & $2(l+1)$ \\
\hline
\hspace{-2mm}$A_\mu$ & $1,3,\cdots,l$ & $2(l+1)$ \\
\hline
$A_{I,+}$ & $1,3,\cdots,l$ & $2(l-1)$\\
\hline
$A_{I,-}$ & $1,3,\cdots,l$ & $2(l+3)$\\
\hline
\end{tabular}
\end{center}
\caption{Degeneracy of states in the case $\hg_2=\hg_3$ and $l\geq 3$ odd.}
\label{tabdis}
\end{table}

\begin{figure} [h]
\begin{center}
\epsfysize=3.5cm\epsfbox{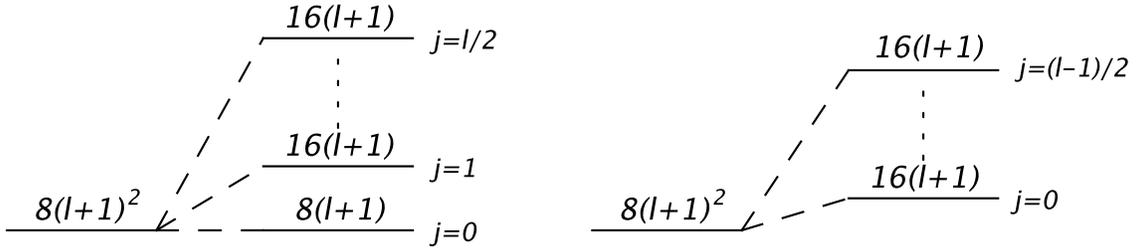}
\end{center}
\caption{The Zeeman--splitting of the undeformed $8(l+1)^2$ d.o.f. for $\hg_2=\hg_3$ and $l$ even 
(left) or odd (right).}
\label{susy}
\end{figure}

\noindent
For any value of $l \geq 2$ we observe Zeeman--like splitting as shown in Fig. \ref{susy}. 
Precisely, the splitting occurs in the following way:
For $l$ even there are $8(l+1)$ d.o.f. corresponding to $j=0$ and $16(l+1)$ for each $j\neq 0$. 
Since we have $l/2$
possible values of $j\neq 0$, the total number of states sum up correctly to $8(l+1)^2$. 
Analogously, for odd values of $l$ the number of levels is $(l+1)/2$, each of them
corresponds to $16(l+1)$ d.o.f., so we still have $8(l+1)^2$ modes.

\noindent
The $l=0$ case corresponds to $m_2=m_3=0$ ($j=0$). The deformation is then harmless and we
are back to the bosonic content of the undeformed theory, that is  three scalars $\chi$, $\Phi$, 
$A_{(I,-)}$ and one vector with $M^{(0)}(n,0)$. 
Similarly, for $l=1$ ($j=0$), excluding $A_{(I,+)}$ we have three scalars and
one vector in the $(1/2,1/2)$ of $SO(4)$ and one scalar in the $(3/2,1/2)$, all 
corresponding to  $M^2 = (M^{(0)}(n,1))^2 + \hg^2 L^2/R^4$. 
These cases can be included in Tables \ref{tabpar} and \ref{tabdis} with the agreement to
discharge modes which are not switched on.

We note that there is an accidental mass degeneracy which is remnant 
of the undeformed $\mathcal{N}=2$ theory. In particular, in the supersymmetric LM case this 
allows to organize the bosonic states in $\mathcal{N}=1$ supermultiplets.

In principle, this unexpected degeneracy could be related 
to the particular theories we are considering which are smooth deformations of their undeformed counterpart. 
In order to better understand $\mathcal{N}=2$ vs. $\mathcal{N}=1$ 
supersymmetry at the level of mesonic spectrum, the study of the fermionic sector is a mandatory requirement.

\vskip 12pt \noindent \underline{$\hg_2\neq\hg_3$}: 
The splitting term $\Delta M$ now depends on both $m_{2,3}$ and no longer on their difference.
In order to make the comparison with the $\hg_2=\hg_3$ case easier, for fixed $l$ it is convenient 
to label $\Delta M$  by two numbers $j$ and $s$ 
\bea\label{nomas}
&&l\quad\mbox{even}\qquad  \Delta M(j,s) = \frac{(j+s)\,\hg_2+(j-s)\,\hg_3}{2} \non\\
&&l\quad\mbox{odd}\qquad  \Delta M(j,s) = \frac{(j+\frac{1}{2}+s)\,\hg_2+(j+\frac{1}{2}-s)\,
\hg_3}{2}
\eea
where $j$ is still defined as before, whereas $s$ is integer if $l$ is even and half--integer if $l$ is odd.
Its range can be read in Tables \ref{tpn} and \ref{tdn}. 

\begin{table}[h]
\begin{center}
\begin{tabular}{||*{4}{c|}|}
\hline
$\mbox{State}$ & $|m_2-m_3|=2j$ & $s$ & $\mbox{Degeneracy}$ \\
\hline
\hline
$\chi,\,\Phi,\,A_{III}$ & 
\hspace{-3.5mm}
\begin{tabular}{c}
$\phantom{.}$\vspace{-2.5mm}\\
$0$\vspace{2.5mm}\\
\hline
$\phantom{aaaa}2,4,\cdots,l\hspace{-7.2mm}\phantom{\int^{P^p}_{P_p}}\phantom{aaaa}$
\end{tabular}
\hspace{-3.5mm}
&
\hspace{-3.5mm}
\begin{tabular}{c}
\hspace{-3.5mm}
\begin{tabular}{c}
$0$\\
\hline
$\phantom{aaaaaaaa}1,2,\cdots,\frac{l}{2}\phantom{aaaaaaaa}$
\end{tabular}
\hspace{-3.5mm}
\\
\hline
$\phantom{aaaa}\hspace{-7.5mm}\phantom{\int^{P^P}_{P_P}}-\frac{l}{2},\cdots,0,\cdots,\frac{l}{2}
\phantom{aaaa}$
\end{tabular}
\hspace{-3.5mm}
&
\hspace{-4mm}
\begin{tabular}{c}
\hspace{-3.5mm}
\begin{tabular}{c}
$\phantom{aaaaaa}1\phantom{aaaaaa}$\\
\hline
$2$
\end{tabular}
\hspace{-3.5mm}
\\
\hline
$\phantom{aaaa}\hspace{-7.5mm}\phantom{\int^{P^P}_{P_P}}2\phantom{aaaa}$
\end{tabular}
\hspace{-6mm}
$\phantom{.}$
\\
\hline
\hspace{-2mm}
$A_{\mu}$ & 
\hspace{-3.5mm}
\begin{tabular}{c}
$\phantom{.}$\vspace{-2.5mm}\\
$0$\vspace{2.5mm}\\
\hline
$\phantom{aaaa}2,4,\cdots,l\hspace{-7.2mm}\phantom{\int^{P^p}_{P_p}}\phantom{aaaa}$
\end{tabular}
\hspace{-3.5mm}
&
\hspace{-3.5mm}
\begin{tabular}{c}
\hspace{-3.5mm}
\begin{tabular}{c}
$0$\\
\hline
$\phantom{aaaaaaaa}1,2,\cdots,\frac{l}{2}\phantom{aaaaaaaa}$
\end{tabular}
\hspace{-3.5mm}
\\
\hline
$\phantom{aaaa}\hspace{-7.5mm}\phantom{\int^{P^P}_{P_P}}-\frac{l}{2},\cdots,0,\cdots,\frac{l}{2}
\phantom{aaaa}$
\end{tabular}
\hspace{-3.5mm}
&
\hspace{-4mm}
\begin{tabular}{c}
\hspace{-3.5mm}
\begin{tabular}{c}
$\phantom{aaaaaa}1\phantom{aaaaaa}$\\
\hline
$2$
\end{tabular}
\hspace{-3.5mm}
\\
\hline
$\phantom{aaaa}\hspace{-7.5mm}\phantom{\int^{P^P}_{P_P}}2\phantom{aaaa}$
\end{tabular}
\hspace{-6mm}
$\phantom{.}$
\\
\hline
$A_{I,+}$ & 
\hspace{-3.5mm}
\begin{tabular}{c}
$\phantom{.}$\vspace{-2.5mm}\\
$0$\vspace{2.5mm}\\
\hline
$\phantom{aaaa}2,4,\cdots,l\hspace{-7.2mm}\phantom{\int^{P^p}_{P_p}}\phantom{aaaa}$
\end{tabular}
\hspace{-3.5mm}
&
\hspace{-3.5mm}
\begin{tabular}{c}
\hspace{-3.5mm}
\begin{tabular}{c}
$0$\\
\hline
$\phantom{a\,aaaaaa}1,2,\cdots,\frac{l-2}{2}\phantom{aaaaaaa\,}$
\end{tabular}
\hspace{-3.5mm}
\\
\hline
$\hspace{-8.9mm}\phantom{\int^{P^P}_{P_P}}-\frac{l-2}{2},\cdots,0,\cdots,\frac{l-2}{2}$
\end{tabular}
\hspace{-3.5mm}
&
\hspace{-4mm}
\begin{tabular}{c}
\hspace{-3.5mm}
\begin{tabular}{c}
$\phantom{aaaaaa}1\phantom{aaaaaa}$\\
\hline
$2$
\end{tabular}
\hspace{-3.5mm}
\\
\hline
$\phantom{aaaa}\hspace{-7.5mm}\phantom{\int^{P^P}_{P_P}}2\phantom{aaaa}$
\end{tabular}
\hspace{-6mm}
$\phantom{.}$
\\
\hline
$A_{I,-}$ & 
\hspace{-3.5mm}
\begin{tabular}{c}
$\phantom{.}$\vspace{-2.5mm}\\
$0$\vspace{2.5mm}\\
\hline
$\phantom{aaaa}2,4,\cdots,l\hspace{-7.2mm}\phantom{\int^{P^p}_{P_p}}\phantom{aaaa}$
\end{tabular}
\hspace{-3.5mm}
&
\hspace{-3.5mm}
\begin{tabular}{c}
\hspace{-3.5mm}
\begin{tabular}{c}
$0$\\
\hline
$\phantom{aaaaaa\,a}1,2,\cdots,\frac{l+2}{2}\phantom{aaaaaa\,a}$
\end{tabular}
\hspace{-3.5mm}
\\
\hline
$\hspace{-9mm}\phantom{\int^{P^P}_{P_P}}-\frac{l+2}{2},\cdots,0,\cdots,\frac{l+2}{2}$
\end{tabular}
\hspace{-3.5mm}
&
\hspace{-4mm}
\begin{tabular}{c}
\hspace{-3.5mm}
\begin{tabular}{c}
$\phantom{aaaaaa}1\phantom{aaaaaa}$\\
\hline
$2$
\end{tabular}
\hspace{-3.5mm}
\\
\hline
$\phantom{aaaa}\hspace{-7.5mm}\phantom{\int^{P^P}_{P_P}}2\phantom{aaaa}$
\end{tabular}
\hspace{-6mm}
$\phantom{.}$
\\
\hline
\end{tabular}
\end{center}
\caption{Degeneracy of states in the case $\hg_2\neq\hg_3$ and $l\geq 2$ even. The degeneracy in the 
fourth column refers to every single pair $(j,s)$.}
\label{tpn}
\end{table}

\begin{table}[h]
\begin{center}
\begin{tabular}{||*{4}{c|}|}
\hline
$\mbox{State}$ & $|m_2-m_3|=2j+1$ & $s$ & $\mbox{Degeneracy}$ \\
\hline
\hline
$\chi,\,\Phi,\,A_{III}$ & $1,3,\cdots,l$ & $-\frac{l}{2},\cdots,\frac{l}{2}$ & $\hspace{-7.5mm}
\phantom{\int^{P^P}_{P_P}}2$ \\
\hline
\hspace{-2mm}$A_\mu$ & $1,3,\cdots,l$ & $-\frac{l}{2},\cdots,\frac{l}{2}$ &  $\hspace{-7.5mm}
\phantom{\int^{P^P}_{P_P}}2$ \\
\hline
$A_{I,+}$ & $1,3,\cdots,l$ & $-\frac{l-2}{2},\cdots,\frac{l-2}{2}$ & $\hspace{-7.5mm}
\phantom{\int^{P^P}_{P_P}}2$\\
\hline
$A_{I,-}$ & $1,3,\cdots,l$ & $-\frac{l+2}{2},\cdots,\frac{l+2}{2}$ & $\hspace{-7.5mm}
\phantom{\int^{P^P}_{P_P}}2$\\
\hline
\end{tabular}
\end{center}
\caption{Degeneracy of states in the case $\hg_2\neq\hg_3$ and $l\geq 3$ odd.}
\label{tdn}
\end{table}

As appears in the Tables the degeneracy is almost completely broken. In fact,
except for the $m_2=m_3=0$ case, only a residual degeneracy $2$ survives due to the fact that the 
mass (\ref{MX}) is invariant under the exchange $(m_2,m_3)\,\rightarrow\,(-m_2,-m_3)$.

\begin{figure} [h]
\begin{center}
\epsfysize=8cm\epsfbox{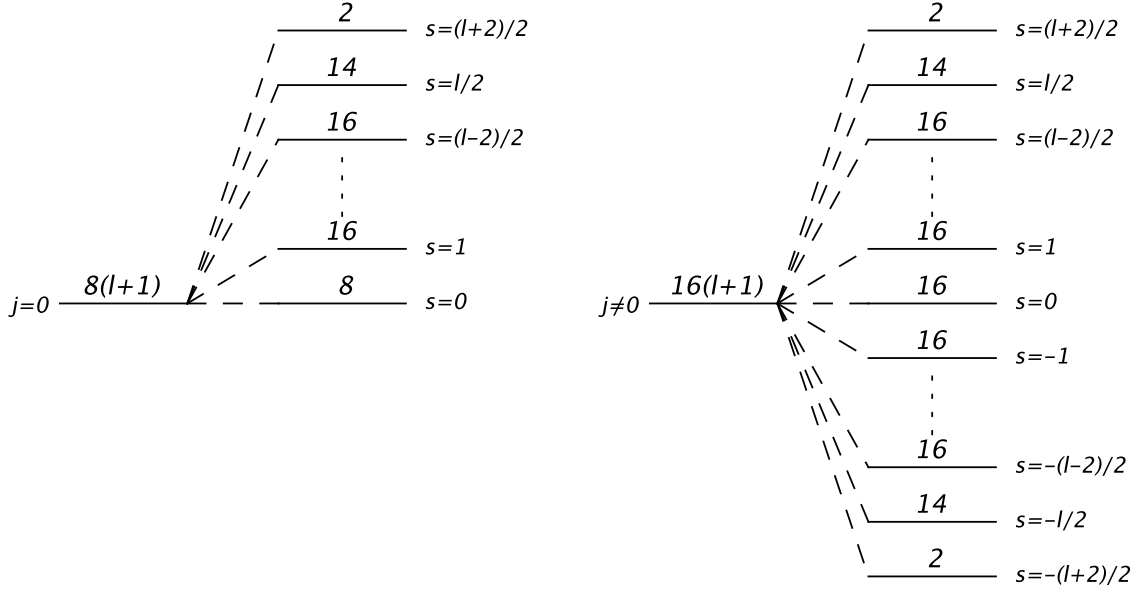}
\end{center}
\caption{The Zeeman--splitting of the $\hg_2=\hg_3=\hg$ d.o.f. for $\hg_2\neq\hg_3$ and  $l$ even. 
The value of $\Delta M$ here appearing is pictured considering the case $\hg_3<\hg<\hg_2$.}
\label{nonpar}
\end{figure}

\begin{figure} [h]
\begin{center}
\epsfysize=8cm\epsfbox{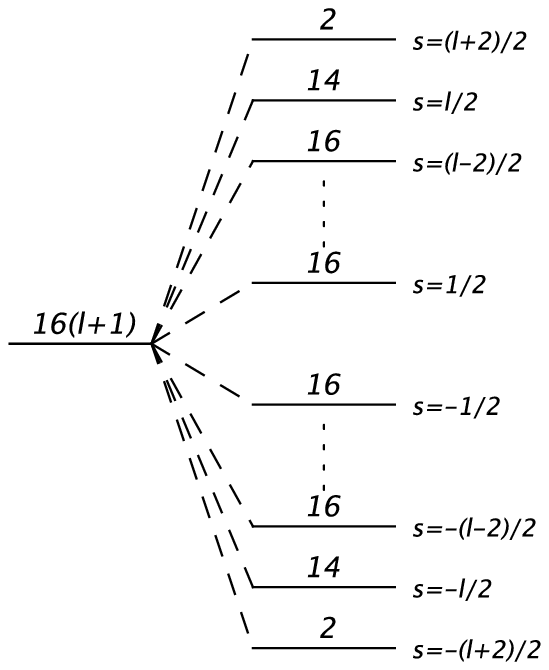}
\end{center}
\caption{The Zeeman--splitting of the $\hg_2=\hg_3$ d.o.f. for $\hg_2\neq\hg_3$ and $l$ odd. 
Once again $\hg_3<\hg<\hg_2$.}
\label{nondis}
\end{figure}

To better understand the level splitting it is convenient to compare the present situation with the
previous one. 
In fact, fixing $j$, the degenerate degrees of freedom of the $\hg_2=\hg_3$ case further 
split according to the different values of $s$.  
If $l$ is even and $j=0$, the previous $8(l+1)$ degenerate levels split in $(l/2+2)$ new mass levels, while 
for $j\neq 0$ the $16(l+1)$ levels open up in $(l+3)$ levels (see Fig. \ref{nonpar}). If $l$ is odd we 
find $(l+3)$ different mass levels as drawn in Fig. \ref{nondis}.

The particular cases $l=0,1$ can be read from Tables \ref{tpn} and \ref{tdn}
by discharging $(A_{(I,+)}, A_{III})$ and $A_{(I,+)}$, respectively. 
For $l=0$ three modes $\chi$, $\Phi$ and $A_\mu$ correspond to $\Delta M=0$ ($j=s=0$), whereas
the three degrees of freedom of $A_{(I,-)}$ split into one d.o.f. with $\Delta M=0$ ($j=s=0$) 
and two with $\Delta M= \frac{\hg_2-\hg_3}{2}$ ($j=s=1$). Already in the simplest $l=0$ case
the $SO(4)$ breaking is manifest.
For $l=1$ ($j=0$) the four degrees of freedom of each mode $\chi$, $\Phi$, $A_{III}$ and 
$A_\mu$ now split into two states with $\Delta M = \hg_2/2$ and two states with
$\Delta M = \hg_3/2$. On the other hand, the $8$ d.o.f. corresponding to $A_{(I,-)}$
split into two states with $\Delta M = \hg_2/2$, two states with 
$\Delta M=\hg_3/2$, two states with $\Delta M = (2\hg_2-\hg_3)/2$
and two with $\Delta M = (2\hg_3-\hg_2)/2$. 

\vskip 15 pt

As discussed in \cite{KMMW} the undeformed spectrum exhibits a huge degeneracy in $\n \equiv n+l$
which can be traced back to a (non--exact) SO(5) symmetry. This originates from the fact that 
the induced metric on the D7--brane is conformally equivalent to $E^{(1,3)} \times {\rm S}^4$. 
If in the quadratic action for the fluctuations the conformal factor can be re--absorbed by a 
field redefinition the corresponding equations of motion are invariant under ${\rm S}^4$ diffeomorphisms.
Therefore, solutions can be found by expanding in spherical harmonics of ${\rm S}^4$  
and the mass spectrum of the elementary modes depends only on the $SO(5)$ quantum number $\n$. 
This happens for instance for scalar modes and vectors which, for a given $\n$,
organize themselves 
into reducible representations $(0,0) \oplus (1/2,1/2) \cdots \oplus (\n/2,\n/2)$ of $SO(4)$. 
This is indeed the decomposition of the highest weight representation $[\n,0]$ of $SO(5)$ 
in $SO(4)$ representations.  

In principle, the same analysis can be applied also to our case. 
Here the induced metric (\ref{metrind}) is conformally
equivalent to $E^{(1,3)} \times \tilde{{\rm S}}^4$ where $\tilde{{\rm S}}^4$ is the deformed four--sphere
(set $\varrho = \rho/L$)
\beq
ds^2_{\tilde{{\rm S}}^4} =  \frac{R^4}{4L^2} \frac{4}{(1+\varrho^2)^2} ( d \varrho^2 + \varrho^2 
d\tilde{\O}_3^2)
\eeq
and
\beq 
d\tilde{\O}_3^2 = d\theta^2 + G \left[c_\theta^2 d\phi_2^2+s_\theta^2 d\phi_3^2+
\frac{\varrho^2  c_\theta^2 s_\theta^2 (\hat{\gamma}_2 d\phi_2+\hat{\gamma}_3 
d\phi_3)^2}{(1+\varrho^2)^2}\right]
\eeq
is the deformed three--sphere.

It follows that a dependence on the $SO(5)$ quantum number $\nu=n+l$ still appears if the conformal factor 
$(1+\varrho^2)L^2/R^2$ can be compensated by a field redefinition and the action can be entirely
expressed in terms of the metric of $E^{(1,3)} \times  {\rm S}^4$ plus deformations. 
A close look at the action (\ref{L2}) reveals that this is always the case for the decoupled 
modes $\chi,\, A_\mu$ and also for $\Phi$. Despite of the presence of the deformation terms 
which break explicitly the $SO(5)$ invariance, we can still
search for solutions expanded in spherical harmonics on ${\rm S}^4$ and, consequently, the mass 
spectrum exhibits a dependence on $n$ and $l$ only in the combination $n+l$. 
In particular, in the zero--mode sector $m_2=m_3=0$ 
a degeneracy appears which is remnant of the $SO(5)$ invariance. 
Of course, the eigenstates corresponding to degenerate eigenvalues never reconstruct 
the complete $[\n,0]$ representation of $SO(5)$, being organized into a direct product of $SO(4)$ 
representations with integer spins only $(0,0) \oplus (1,1) 
\cdots (\left[\n/2\right],\left[\n/2\right])$, since $m_2=m_3=0$ only occurs for even values of $l$.

\section{The dual field theory} 

In this Section we construct the 4D conformal field theory whose composite operators 
are dual to the mesonic states just found. 

As already discussed in Section 3, in the supergravity description the operations of $TsT$ deforming 
the ${\rm AdS}_5 \times {\rm S}^5$ background and adding D7--branes commute. Since on the field theory
side $TsT$ deformations correspond to promoting all the products among the fields to be $\ast$--products 
\cite{LM}, whereas the addition of D7--branes corresponds to adding interacting fundamental matter \cite{KK}
we expect that in determining the action for the dual field theory the operations of $\ast$--product 
deformation and addition of fundamental matter commute. Therefore, in order to obtain the dual action we proceed
by promoting to $\ast$--products all the products in the ${\cal N}=2$ SYM action with fundamental matter
corresponding to the undeformed Karch--Katz model.

Given $N_f$ probe D7--branes embedded in the ordinary ${\rm AdS}_5 \times {\rm S}^5$ background 
with $N$ units of flux, $N \gg N_f$, in the large $N$ limit the dual field theory on the D3--branes consists of 
${\cal N}=4$ $SU(N)$ SYM coupled in a ${\cal N}=2$ fashion to
$N_f$ $\mathcal{N}=2$ hypermultiplets which contain  
new dynamical fields arising from open strings stretching between D3 and 
D7--branes.
In ${\cal N}=1$ superspace language the $\mathcal{N}=4$ gauge 
multiplet is given in terms of one $\mathcal{N}=1$ gauge superfield 
$W_\alpha$ and three chirals $\F_1,\,\F_2,\,\F_3$ all in the adjoint 
representation of $SU(N)$. The ${\cal N}=2$ hypermultiplets are described
by $N_f$ chiral superfields $Q^r$ transforming in the $(N, \bar{N}_f)$
of $SU(N) \times SU(N_f)$ plus $N_f$ chirals $\tilde{Q}_r$ transforming  
in the $(\bar{N}, N_f)$. 

According to the AdS/CFT duality the lowest components of the three chirals 
$\F_i$ are in one--to--one
correspondence with the three complex coordinates of the internal 6D space as 
(we use notations consistent with Section 2)
\bea\label{comcoord}
&& X^1 + i X^{2} \equiv u \rho_3 e^{i\phi_3} \to \F_3|_{\th = \bar{\th} =0} \non \\   
&& X^3 + i X^{4} \equiv u \rho_2 e^{i\phi_2} \to \F_2|_{\th = \bar{\th} =0} \\   
&& X^5 + i X^{6} \equiv u \rho_1 e^{i\phi_1} \to \F_1|_{\th = \bar{\th} =0}\non
\eea
For a configuration of D7--branes placed at distance $X^5+iX^6=L$ from the 
D3--branes the Lagrangian of the corresponding gauge theory is \cite{KK}
\bea \label{lagr1} 
\mathcal{L}&=&\int\ \!d^4\theta\left[ 
  {\rm Tr}\left(e^{-g\,V}\bar{\Phi}_ie^{g\,V} \Phi^i\right)+ 
  {\rm tr} \left(\bar{Q}e^{g\,V}Q+\tilde{Q}e^{-g\,V}\bar{\tilde{Q}}\right)\right]
+\frac{1}{2 g^2}\int\ \! d^2\theta\, {\rm Tr}\left(W^{\alpha}W_{\alpha}\right)
\nonumber\\ 
&+&i\int \!d^2\theta \left[g\,{\rm Tr} \left(\Phi^1\left[\Phi^2,\Phi^3\right]\right)+ 
  g\,{\rm tr} \!\left(\tilde{Q}\Phi^1 Q\right)+ m\, {\rm tr}\left(\tilde{Q}Q\right)
  \right]+h.c.
\eea
where the trace Tr is over color indices and tr is over the flavor
ones. This action is ${\cal N}=2$ supersymmetric with
$(W_\a, \F_1)$ realizing a $\mathcal{N}=2$ vector multiplet and $(\F_2,\F_3)$
an adjoint matter hypermultiplet. 
The coupling of $\F_1$ with massive matter fields leads to a non--trivial
vev $\langle \F_1 \rangle = -m/g$ which gives the displacement between the D3 and the 
D7--branes according to the identification $L \equiv -m/g$.     

The theory has a $SU(2)_{\F} \times SU(2)_R$ invariance corresponding to a symmetry 
which exchanges $(\F_2, \F_3)$ and to the ${\cal N}=2$ R--symmetry, respectively.
In addition, for $m=0$, there is a $U(1)$ R--symmetry under which $(Q^r,\tilde{Q}_r)$
and $(\F_2, \F_3)$ are neutral, whereas $\F_1$ has charge $2$ and $W_\a$ has charge $1$
\cite{HYS,kirsch}. 
In the dual supergravity description these symmetries originate from the $SO(4)\times SO(2)$ 
invariance which survives after the insertion of the D7--branes \cite{KK} and which are related 
to rotations in the $(X^1, X^2,X^3,X^4)$ and $(X^5,X^6)$ planes, respectively.       
Fixing $X^5+iX^6=L \neq 0$ breaks rotational invariance in the $(X^5,X^6)$ plane
and, correspondingly, the mass term breaks the $U(1)$ R--symmetry in the dual gauge theory. 
Finally, the theory also possesses a $U(1)$ baryonic symmetry under which only $(Q^r,\tilde{Q}_r)$
are charged $(1,-1)$. This is a residual of the original $U(N_f)$ invariance.

For $m=0$ and in the large $N$ limit with $N_f$ fixed the theory is superconformal invariant. In fact, 
the beta--function for the 't Hooft coupling $\l = g^2 N$ is proportional to $\l^2 N_f/N$
and vanishes for $N_f/N \to 0$.  

Since we are interested in non--supersymmetric deformations of this theory
we need the Lagrangian (\ref{lagr1}) expanded in components. Given the physical components
of the multiplets being 
\bea
\Phi^i&=&\left(a^i,\psi^i_\a\right) \qquad \qquad ~Q^r = \left(q^r,  \chi^r_\a \right)\non\\
W_\a &=& (\l_\a , f_{\a\b}) \qquad \qquad \tilde Q_r = \left(\tilde{q}_r,\tilde{\chi}_{r \a}\right)
\eea
after eliminating the auxiliary fields through their algebraic equations
of motion, the Lagrangian (\ref{lagr1}) takes the form
\beq
\mc{L}=\mc{L}_{\mc{N}=4}+\mc{L}_{b}+\mc{L}_{f}+\mc{L}_{int}
\label{lagr2}
\eeq
where \footnote{We use superspace conventions of \cite{superspace}. When $\psi \l$
indicates the product of two chiral fermions it has to be understood as $\psi^\a \l_\a$.
The same convention is used for antichiral fermions.}  
\bea
\mathcal{L}_{\mc{N}=4}&=& {\rm Tr} \left( -\frac{1}{2}f^{\alpha\beta}
f_{\alpha\beta}+
i\lambda\left[\nabla,\bar{\lambda}\right]+ \bar{a_i}\Box a^i+
i{\psi}^i\left[\nabla,\bar{\psi}_i\right] \right) \non \\
&~~&+g^2\ \!{\rm Tr}\left(-\frac{1}{4}\left[a^i,\bar{a}_i\right]\left[a^j,
\bar{a}_j
\right]+\frac{1}{2}\left[a^i,a^j\right]\left[\bar{a}_i,\bar{a}_j
\right]\right)\nonumber \\
&~~&+\Bigg\{i g \,{\rm Tr}\left(\left[\bar{\psi}_i,\bar{\lambda}\right]a^i
+\frac{1}{2}\e_{ijk}\left[\psi^i,\psi^j\right] a^k\right)\ \  +h.c.\Bigg\}
\eea
is the ordinary ${\cal N}=4$ Lagrangian,
\bea
\mc{L}_{b}&=&{\rm tr}\Big(\bar{q}\left(\Box -|m|^2\right)q
+\tilde{q}\left(\Box-|m|^2\right) \bar{\tilde{q}}\Big) \non\\
&-&\frac{g^2}{4}\ \!{\rm tr} \bigg(\bar{q}\,q\,\bar{q}\,q+
\tilde{q}\,\bar{\tilde{q}}\,\tilde{q}\,\bar{\tilde{q}}
-2\bar{q}\,\bar{\tilde{q}}\,\tilde{q}\,q
+4\tilde{q}\,\bar{\tilde{q}}\,\bar{q}\,q\bigg)+ 
\frac{g^2}{2}\ \! {\rm tr} \left(\tilde{q}\left[a^i,\bar{a}_i\right]
\bar{\tilde{q}}-\bar{q}\left[a^i,\bar{a}_i\right]q
\right) \non \\
&-&\Bigg\{{\rm tr}\left(g \bar{m}\,\!\!\left(\bar{q}a_1q+\,\tilde{q}a_1\bar{\tilde{q}}
\right)
+\frac{g^2}{2}\,\!\!\left(\bar{q}\bar{a_1}a^1q+\tilde{q}a^1\bar{a_1}
\bar{\tilde{q}}+2 \tilde{q}
\left[\bar{a}_2,\bar{a}_3\right]q\right)\right)\ \ +h.c.\Bigg\}
\eea
describes the bosonic fundamental sector and its interactions with bosonic matter in the
adjoint,
\beq
\mc{L}_{f}=i\ \!{\rm tr} \left(\bar{\chi}\overrightarrow{\nabla}\chi-\tilde{\chi}
\overleftarrow{\nabla}\bar{\tilde{\chi}}\right)
+ \Big\{i m\ \! {\rm tr} \bigg(\tilde{\chi}\chi\bigg)\ \ +h.c.\Big\}
\eeq
describes the free fermionic fundamental sector and 
\beq
\mc{L}_{int}= i g\ \!{\rm tr}\bigg(\bar{\chi}\bar{\lambda}q
    -\tilde{q}\bar{\lambda}\bar{\tilde{\chi}}
+\tilde{q}\psi^1\chi+\tilde{\chi}\psi^1 q+
\tilde{\chi}a^1\chi\bigg)\ \ +h.c.
\eeq
contains the interaction terms between bosons and fermions.

The most general non--supersymmetric marginal deformation of this theory can be 
obtained by promoting all the products among the fields in the Lagrangian to be
$\ast$--products according to the following prescription \cite{BRFT}
\beq
f\,g \,\,\longrightarrow\,\,f \ast g=e^{i\pi Q^f_i Q^g_j \e_{ijk}\gamma_k}\,f\,g
\label{star}
\eeq
where $\g_k$ are the deformation parameters, whereas
$(Q_1,Q_2,Q_3)$ are the charges of the fields under the three $U(1)$
global symmetries of the original ${\cal N}=4$ theory associated to the Cartan generators 
of $SU(4)$. On the dual supergravity side they correspond to angular shifts in 
(\ref{comcoord}). Accordingly, the charges of the chiral $\F_i$ superfields are 
chosen as in Table \ref{Tableaux} \cite{BRFT} with the additional requirement for the charges of the 
spinorial superspace coordinates to be $(1/2,1/2,1/2)$. This insures invariance of
the superpotential term $\int d^2\th {\rm Tr} (\Phi^1[\Phi^2,\Phi^3])$
under the three $U(1)$'s. The charges for the matter chiral superfields 
are determined by requiring the superpotential term
$\int d^2\th {\rm tr}(\tilde{Q}\Phi^1 Q)$ to respect the three global symmetries
in addition to the condition for $Q$ and $\tilde{Q}$ to have the same charges.  

\begin{table}[h]
\begin{center}
\begin{tabular}{|c||c|c|c|c|c|}
  \hline
  & $\Phi^1$ & $\Phi^2$ & $\Phi^3$ & $Q$ & $\tilde{Q}$\hspace{-7mm}$\phantom{\int^{F^{F}}}$\\
  \hline
\hline
  $Q_1$ & $1$ & $0$ & $0$ & $0$ & $0$\\
  \hline
  $Q_2$ & $0$ & $1$ & $0$ & $\frac12$ & $\frac12$\\
  \hline
  $Q_3$ & $0$ & $0$ & $1$ & $\frac12$ & $\frac12$\\
  \hline
\end{tabular}
\caption{$U(1)$ charges of the chiral superfields. The corresponding antichirals have opposite charges.}
\label{Tableaux}
\end{center}
\end{table}
The gauge superfield $W_\a$ and the gaugino have charges $(1/2,1/2,1/2)$, whereas 
the gauge field strength $f_{\alpha\beta}$ is neutral under the three $U(1)$'s.

In the absence of mass term in (\ref{lagr1}) 
the corresponding currents $(J_{\phi_1},J_{\phi_2},J_{\phi_3})$ are conserved,
whereas $J_{\phi_1}$ fails to be conserved when $m \neq 0$.
Moreover, $(J_{\phi_2},J_{\phi_3})$ are ABJ--anomaly
free also in the presence of fundamental matter, whereas $J_{\phi_1}$ is 
non--anomalous only in the quenching limit $N_f/N \to 0$.

As is well--known, the ordinary Lunin--Maldacena $U(1) \times U(1)$ charges \cite{LM}
are associated to $(\varphi_1, \varphi_2)$ angular shifts after performing the change of 
variables (in our notations)
\beq 
\varphi_1 = \frac13(\phi_1 + \phi_2 - 2\phi_3),  \quad
\varphi_2 = \frac13(\phi_2 + \phi_3 - 2\phi_1), \quad
\varphi_3 = \frac13(\phi_1 + \phi_2 + \phi_3), 
\eeq 
Expressing the $(J_{\varphi_1}, J_{\varphi_2})$ generators in terms of $(J_{\phi_1}, J_{\phi_2},
J_{\phi_3})$ we easily find that the Lunin--Maldacena charges are given by
\beq \label{relazione}
Q^{(LM)}_1 = Q_2-Q_3 \qquad , \qquad Q^{(LM)}_2 = Q_2-Q_1
\eeq
In the case of supersymmetric deformations the third linear combination 
$Q_R \sim (Q_1 + Q_2 + Q_3)$ provides the R--symmetry charge. 

We are now ready to derive the deformed action by using the prescription (\ref{star}) in 
the original undeformed one. 

We begin with the one--parameter deformation, $\g_1=\g_2=\g_3$.
In this case ${\cal N}=1$ supersymmetry survives and we can work directly with the superspace
action (\ref{lagr1}). Since only for $m=0$ the $\ast$--product is well--defined being the three U(1) 
charges conserved, the correct way to proceed is to deform the massless theory and then add
the mass operator as a perturbation. Following this prescription and taking into account the superfields
charges given in Table \ref{Tableaux}, the Lagrangian of the deformed theory is 
\bea
\mathcal{L}&=&\int\ \!d^4\theta\left[ 
  {\rm Tr}\left(e^{-g\,V}\bar{\Phi}_ie^{g\,V} \Phi^i\right)+ 
  {\rm tr} \left(\bar{Q}e^{g\,V}Q+\tilde{Q}e^{-g\,V}\bar{\tilde{Q}}\right)\right]
+\frac{1}{2 g^2}\int\ \! d^2\theta\, {\rm Tr}\left(W^{\alpha}W_{\alpha}\right)
\nonumber\\ 
&+&  i g \int d^2\th \, \left[
\mbox{Tr}\left(e^{i \pi \gamma} \Phi_1 \Phi_2 \Phi_3-e^{-i \pi \gamma} \Phi_1 
\Phi_3 \Phi_2\right) + \,\mbox{tr} \left(\tilde{Q} \Phi_1 Q\right) 
+  m \,\mbox{tr} \left(\tilde{Q} Q\right) \right]
\label{Wdef}
\eea
We note that a non--trivial deformation appears in the superpotential 
only in the pure adjoint sector. The interaction and the mass terms involving 
flavor matter do not change, so that the vev for $\F_1$ which is related
to the D7--brane location through the dictionary (\ref{comcoord}) is the same as 
in the undeformed theory, $ \langle \F_1 \rangle = -m/g \equiv L$. 
Since in the supergravity description we
have chosen $L$ to be real ($X^5=L$, $X^6=0$) here and in what follows we restrict to real 
values of $m$.  

As already stressed, for $m \neq 0$ the $Q_1$ charge is not conserved, neither is $Q^{(LM)}_2$. Therefore, 
this deformed theory possesses only one U(1) non--R--symmetry corresponding to $Q^{(LM)}_1$. 

The action (\ref{Wdef}) has been obtained by $\ast$--product deforming the ${\cal N}=2$ SYM action
(\ref{lagr1}). However, it could have been equivalently obtained
by adding fundamental chiral matter to the ${\cal N}=1$ $\b$--deformed SYM theory of \cite{LM}. 
In particular, the appearance of the gauge coupling constant in front of the adjoint chiral 
superpotential insures that for $m=0$ and in the probe approximation the theory is superconformal invariant 
\cite{MPSZ}.

\vskip 15pt
We now consider the more general non--supersymmetric case. We implement the
$\ast$--product (\ref{star}) in the action (\ref{lagr2}). 
Using the deformed commutator \cite{BRFT}
\begin{equation}
[X_i,X_j]_{M_{ij}} \equiv e^{i \pi M_{ij}} X_iX_j-e^{-i \pi M_{ij}} X_jX_i
\end{equation}
where for $X_i$ fermions 
\bea
M_{\rm fermions} \equiv B =\left(\begin{array}{cccc}
0&\frac{1}{2}(\gamma_1+\gamma_2)
&-\frac{1}{2}(\gamma_1+\gamma_3)&-\frac{1}{2}(\gamma_2-\gamma_3)\\
-\frac{1}{2}(\gamma_1+\gamma_2)&0
&\frac{1}{2}(\gamma_2+\gamma_3)&-\frac{1}{2}(\gamma_3-\gamma_1)\\
\frac{1}{2}(\gamma_3+\gamma_1)&-\frac{1}{2}(\gamma_2+\gamma_3)
&0&-\frac{1}{2}(\gamma_1-\gamma_2)\\
\frac{1}{2}(\gamma_2-\gamma_3)&\frac{1}{2}(\gamma_3-\gamma_1)&
\frac{1}{2}(\gamma_1-\gamma_2)&0\\
\end{array}
\right)&
\eea
whereas for scalars
\bea
M_{\rm scalars} \equiv C =\left(\begin{array}{ccc}
0&\gamma_3&-\gamma_2\\
-\gamma_3&0&\gamma_1\\
\gamma_2&-\gamma_1&0\\
\end{array}\right)
&
\eea
the deformed $\mc{L}_{\mc{N}=4}$ takes the form
\bea
\mathcal{L}_{\mathcal{N}=4}&=& {\rm Tr} \left( -\frac12 f^{\alpha\beta}f_{\alpha\beta}+
i\lambda\left[\nabla,\bar{\lambda}\right]+ \bar{a_i}\Box a^i+
i{\psi}^i\left[\nabla,\bar{\psi}_i\right] \right) \non \\
&+&g^2 \, {\rm Tr} \left(-\frac{1}{4}\left[a^i,\bar{a}_i\right]\left[a^j,\bar{a}_j
  \right]+\frac{1}{2}\left[a^i,a^j\right]_{C_{ij}}\left[\bar{a}_i,
  \bar{a}_j \right]_{C_{ij}}\right)\nonumber \\
&+& \Bigg\{i g \, {\rm Tr} \left( \left[\bar{\psi}_i,\bar{\lambda}\right]_{B_{i4}}a^i
+\frac{1}{2}\e_{ijk}\left[\psi^i,\psi^j\right]_{B_{ij}}a^k\right)\ \ +h.c.\Bigg\}
\eea 
while the bosonic sector reads
\bea
\mc{L}_{b}&=&{\rm tr}\Big(\bar{q}\left(\Box -m^2\right)q
+\tilde{q}\left(\Box-m^2\right) \bar{\tilde{q}}\Big) 
-\frac{g^2}{4}\ \!{\rm tr} \bigg(\bar{q}\,q\,\bar{q}\,q+
\tilde{q}\,\bar{\tilde{q}}\,\tilde{q}\,\bar{\tilde{q}}
-2\bar{q}\,\bar{\tilde{q}}\,\tilde{q}\,q
+4\tilde{q}\,\bar{\tilde{q}}\,\bar{q}\,q\bigg) \non \\
&+& \frac{g^2}{2}\ \!{\rm tr}\left(\tilde{q}\left[a^i,\bar{a}_i\right]
\bar{\tilde{q}}-\bar{q}\left[a^i,\bar{a}_i\right]q +
\bar{q}\bar{a_1}a^1q+\tilde{q}a^1\bar{a_1}\bar{\tilde{q}}\right)
\ \!\!\  \non \\
&+& \Big\{ g^2\ \! {\rm tr}\left(\tilde{q}\left[\bar{a}_2,\bar{a}_3\right]_{C_{23}}q \right)
- gm \, {\rm tr} \left(e^{-i \pi \left(\g_2-\gamma_3\right)}\bar{q}a^1q+
\,e^{i \pi  \left(\g_2-\g_3\right)}
\tilde{q}a^1\bar{\tilde{q}} \right) 
\ \!\!\ +h.c. \Big\}\non \\
\eea
and the fermionic one
\beq
\mc{L}_{f}= i\ \!{\rm tr}\left(\bar{\chi}\overrightarrow{\nabla}\chi-\tilde{\chi}
\overleftarrow{\nabla}\bar{\tilde{\chi}}\right)
+ \Big\{im\ \! {\rm tr}\bigg(\tilde{\chi}\chi\bigg)\ \ +h.c.\Big\}
\eeq
Finally the boson--fermion interaction terms become
\bea
\mc{L}_{int}&=& i g\ \! {\rm tr}\bigg(
e^{i \frac{\pi }{4}\left(\g_2-\g_3\right)}\bar{\chi}\bar{\lambda}q
-e^{-i \frac{\pi}{4} \left(\g_2-\g_3\right)}
\tilde{q}\bar{\lambda}\bar{\tilde{\chi}}
\nonumber \\
&&\phantom{carlo\!\!}+
e^{i\frac{\pi}{4} \left(\g_2-\g_3\right)}\tilde{q}\psi^1\chi
+e^{-i\frac{\pi}{4}\left(\g_2-\g_3\right)}\tilde{\chi}\psi^1q+
\tilde{\chi}a^1\chi\bigg)\ \ \!\!+h.c.
\eea
We observe that the fundamental fields $q$ and $\tilde{q}$ experiment the
$\g_1$--deformation only through the modified commutator $\left[\bar{a}_2,
\bar{a}_3\right]_{C_{23}}$ in $\mathcal{L}_b$.  Moreover, $\g_2$ and
$\g_3$ are always present in the combination $(\g_2-\g_3)$ so that the corresponding phases 
disappear when $\gamma_2=\gamma_3$, in particular for supersymmetric deformations.

\section{Conclusions}

In this paper we have studied the embedding of D7--branes in LM--Frolov
backgrounds with the aim of finding the mesonic spectrum of the dual Yang--Mills theory 
with flavors, according to the gauge/gravity correspondence.
Since these theories have ${\cal N}=1$ or no supersymmetry depending on the choice of the deformation 
parameters $\hat{\g}_i$, they provide an interesting playground
in the study of generalizations of the AdS/CFT correspondence to more realistic models with
less supersymmetry.

These geometries are smoothly related to the standard ${\rm AdS}_5\times {\rm S}^5$ from which they can
be obtained by operating with $TsT$ transformations. Therefore, 
if we consider D7--brane embeddings which closely mimic the ones of the undeformed case \cite{KK}
we expect the flavor probes to share some properties with the probes of
the undeformed case. Driven by this observation 
we have considered a spacetime filling D7--brane wrapped on a deformed three--sphere in the 
internal coordinates. We have found that for both the supersymmetric and
the non--supersymmetric deformations a static configuration exists which is 
completely independent of the specific values of the
deformation parameters $\hat{\gamma}_i$. As a consequence the D7--brane still
lies at fixed values of its transverse directions and exhibits no quark
condensate \cite{KK}. We remark that this shape is exact and
stable in the supersymmetric as well as in the non--supersymmetric cases.

Although the shape of the brane does not feel the effects of the
deformation, its fluctuations do. In fact, studying the scalar and vector fluctuations 
we have found that a non--trivial dependence on the $\hat{\gamma}_{2,3}$ parameters
appears both in terms which correct the free dynamics of the modes and in terms which couple
the $U(1)$ worldvolume gauge field to one of the
scalars in the mutual orthogonal directions to the D3--D7 system.
All the deformation--dependent contributions arise from the Dirac--Born--Infeld term in the
D7--brane action, whereas the Wess--Zumino term does not feel the deformation. 
The $\hat{\gamma}_1$ parameter, associated to a $TsT$ transformation along the torus inside 
the D7 worldvolume, never enters the equations of motion.

A smooth limit to the undeformed equations of motion exists
for $\hat{\gamma}_i \rightarrow 0$.
In this limit all the modes decouple and we are back to the undeformed solutions of
\cite{KMMW}. The effect of the deformations becomes negligible also in the 
UV limit ($\rho \to \infty$). This is an expected
result since the deformations involve tori in the internal space and in the UV limit
the metric of the background reduces to flat four dimensional
Minkowski spacetime.

On the other hand, the situation changes once we consider
the general deformed equations. In fact, solving analytically these equations for
elementary excitations of scalars and vectors we have found that  
the mass spectrum is still discrete and with a mass gap and the corresponding
eigenstates match the one of the undeformed case. However, the
mass eigenvalues acquire a non--trivial dependence on $\hat{\gamma}_{2,3}$. 
These new terms, being proportional to the $U(1) \times U(1)$ quantum numbers $(m_2,m_3)$,
induce a level spitting according to a Zeeman--like effect. 

We have performed a detailed analysis of the level splitting and of the corresponding
degeneracy.
The situation turns out to be very different according to $\hg_2$ and $\hg_3$ being
equal or not. In fact, for $\hg_2 \neq \hg_3$ the degeneracy is almost completely broken since only
a residual degeneracy associated to the invariance of the mass under $(m_2,m_3) \to
(-m_2, -m_3)$ survives. In particular, the breaking of $SO(4)$ is manifest. 
Instead, for $\hg_2=\hg_3$   
the mass levels split but for each value of the mass an accidental degeneracy survives
which is remnant of the ${\cal N}=2$ case. While in the supersymmetric case ($\hg_1=\hg_2=\hg_3$)
this allows to arrange mesons in massive $\mathcal{N}=1$ multiplets according to the
fact that our embedding preserves supersymmetry, this higher degree of degeneracy in the 
bosonic sector of the theory does not have a clear explanation at the moment.  
In order to make definite statements about the supersymmetry properties of the mesonic spectrum
and supersymmetry breaking one should study the fermionic sector. A useful strategy could be the 
bottom--up approach described in \cite{kirsch}. We leave this interesting open problem for the future.

Our analysis shares some similarities with other cases considered in the
literature. 

First of all, we have found that a stable embedding of 
the probe brane can be realized which is static and independent of the 
deformation parameters.
This feature has been already encountered for other brane configurations in 
deformed backgrounds. An example is given by particular dynamical probe D3--branes (giant
gravitons) which have been first well understood in \cite{GODB}. In
fact, there it has been shown that giant gravitons exist and are stable even
in the absence of supersymmetry and their dynamics turns
out to be completely independent of the deformation parameters, being then equal
to the one of the undeformed theory.  
Moreover, since the giants wrap the same cycle inside the internal deformed space
as our D7--brane does, their bosonic fluctuations encode the same dependence on the deformation
parameters observed in the mesonic spectrum coming from the D7.

A second similarity emerges with the case of flavors in non--commutative
theories investigated in \cite{NC}. In fact,
the non--trivial coupling between scalar and gauge modes that in our case is induced by the 
deformation resembles the one which appears in the case of
D7--branes embedded in ${\rm AdS}_5 \times {\rm S}^5$ with a $B$ field
turned on along spacetime directions. This is not surprising since both theories can be
obtained performing a $TsT$ transformation of ${\rm AdS}_5\times {\rm S}^5$: 
If the $TsT$ is performed in AdS one obtains the dual of a
non--commutative theory while the LM--Frolov picture is recovered if
this transformation deforms the internal ${\rm S}^5$. 

The field theory dual to the (super)gravity picture we have considered 
can be obtained by deforming the standard action for
${\cal N}=4$ super Yang--Mills coupled to massive $\mathcal{N}=2$ hypermultiplets
by the $\ast$--product prescription \cite{LM}. In principle, in the supergravity dual description
this should correspond to performing a $TsT$ deformation {\em after} the embedding of the probe brane. 
However, as we have discussed, adding the flavor 
brane in the deformed background or deforming the Karch--Katz D3--D7 configuration 
are commuting operations. Therefore, the prescription we propose on the field theory side
is consistent with what we have done on the string theory side. It is important to stress
that the choice of the embedding we have made is crucial for the above reasoning. 

What we obtain is a deformed gauge field theory with massive fundamental matter
parametrized by four real parameters $\g_i$ and $m$.  
We can play with them in order to break global $U(1)$ symmetries, conformality and/or 
supersymmetry in a very controlled way. 
In fact, in the quenching approximation a non--vanishing mass parameter related to 
the location of the probe in the dual geometry breaks conformal invariance and one of the
$U(1)$ global symmetries of the massless theory. 
On the other hand, the values of the deformation parameters $\gamma_i$ determine the degree
of supersymmetry of the theory, as already discussed. 
It is interesting to note that as we found on the gravity side, the three deformation
parameters play different roles in the fundamental sector of the theory. In
fact, $\gamma_{2,3}$ always appear in the combination
$(\gamma_2-\gamma_3)$, so that if $\gamma_2=\gamma_3$ this sector gets deformed only by
$\gamma_1$--dependent phases induced by the interaction with the adjoint matter. 
In the supersymmetric case this particular behavior is manifest
when using superspace formalism since a non--trivial deformation appears only in
the adjoint sector, whereas the flavor superpotential remains undeformed.

Let us conclude mentioning some directions in which our work could be
extended. We have considered only the non--interacting mesonic
sector. Expanding the D7--brane action beyond the second order in
$\alpha^\prime$ one can get informations on the interactions among the
mesons and understand how the deformation enters the
couplings. Moreover, one could extend our analysis to mesons with
large spin in Minkowski, similarly to what has been done in the ordinary, undeformed
case \cite{KMMW}.

Finally it could be very interesting to study in detail the
other embeddings proposed in \cite{mariotti} and in particular the one
which seems to exhibit chiral symmetry breaking. Moreover, going beyond
the quenching approximation has been representing an interesting subject since the
recent efforts to study back--reacted models \cite{back}.

\section*{Acknowledgments}
\noindent 
S.P. thanks M. Grisaru and M.P. thanks A. Butti, D. Forcella 
and A. Mariotti for useful conversations. 
This work has been supported in part by INFN, PRIN prot. 2005024045-002
and the European Commission RTN program MRTN--CT--2004--005104.

\end{document}